\documentclass[trackchanges]{aastex7}

\usepackage{physymb}
\usepackage{mathrsfs}
\usepackage{soul}
\usepackage{color}

\def\fexx{Fe\,{\sc xx}}

\def\fexxiv{Fe\,{\sc xxiv}}
\def\fexxv{Fe\,{\sc xxv}}
\def\fexxvi{Fe\,{\sc xxvi}}

\def\mathv{\textbf{\em v}}

\def\cm{\ifmmode {\rm cm}^{-1} \else cm$^{-1}$ \fi}
\def\s{\ifmmode {\rm s}^{-1} \else s$^{-1}$ \fi}
\def\cc{\ifmmode {\rm cm}^{-3} \else cm$^{-3}$ \fi}
\def\cs{\ifmmode {\rm cm}^{-2} \else cm$^{-2}$ \fi}
\def\g{\ifmmode \gamma \else $\gamma$\fi}
\def\G{\ifmmode \Gamma \else $\Gamma$\fi}
\def\Gs{\ifmmode \Gamma~ \else $\Gamma~$\fi}

\def\gc{\ifmmode \gamma_{\rm c} \else $\gamma_{\rm c}$ \fi}
\def\sw{Schwarzschild~}
\def\gsim{\mathrel{\raise.5ex\hbox{$>$}\mkern-14mu
             \lower0.6ex\hbox{$\sim$}}}
\def\lsim{\mathrel{\raise.3ex\hbox{$<$}\mkern-14mu
             \lower0.6ex\hbox{$\sim$}}}
\def\simless{\mathbin{\lower 3pt\hbox
     {$\rlap{\raise 5pt\hbox{$\char'074$}}\mathchar"7218$}}}   
\def\simmore{\mathbin{\lower 3pt\hbox
     {$\rlap{\raise 5pt\hbox{$\char'076$}}\mathchar"7218$}}}   
\def\Msun{M_\odot}                                
\def\deg{^\circ}

\def\mathv{\textbf{\em v}}

\def\mathu{\textbf{\em u}}

\def\gro1655{GRO~J1655-40}
\def\4u1630{4U1630-472}
\def\h1743{H1743-322}
\def\grs1915{GRS1915+105}

\begin{document}

\title{Implications of Non-equatorial Relativistic Accretion Flows for Ultra-Fast Inflows in AGNs}

\author[orcid=0000-0001-5709-7606]{Keigo Fukumura}
\affiliation{Department of Physics and Astronomy, James Madison University, Harrisonburg, VA 22807, USA}
\email[show]{fukumukx@jmu.edu}  

\author[orcid=0000-0003-2196-3298]{Alessandro Peca}
\affiliation{Eureka Scientific, 2452 Delmer Street, Suite 100, Oakland, CA 94602-3017, USA}
\affiliation{Department of Physics, Yale University, P.O. Box 208120, New Haven, CT 06520, USA}
\email[show]{peca.alessandro@gmail.com}  

\author[orcid=0000-0003-1200-5071]{Roberto Serafinelli}
\affiliation{Instituto de Estudios Astrof\'{i}sicos, Facultad de Ingenier\'{i}a y Ciencias, Universidad Diego Portales, Av. Ej\'{e}rcito Libertador 441, Santiago, Chile}
\affiliation{INAF - Osservatorio Astronomico di Roma, Via Frascati 33, 00078, Monte Porzio Catone (Roma), Italy}
\email[show]{roberto.serafinelli@mail.udp.cl}  

\author[orcid=0000-0002-7858-7564]{Mauro Dadina}
\affiliation{INAF - Osservatorio di Astrofisica e Scienza dello Spazio (OAS) di Bologna, via P. Gobetti 93/3, I-40129 Bologna, Italy}
\email[show]{mauro.dadina@inaf.it}

\begin{abstract}

Motivated by a number of X-ray observations of active galactic nuclei (AGNs) that exhibit a potential signature of ultra-fast inflows (UFIs), we consider in this work a scenario that UFIs can be physically identified as weakly-magnetized hydrodynamic accretion flows that is guided and channeled by poloidal magnetic field into low-to-mid latitude above the  equatorial disk. In the context of general relativistic hydrodynamics (GRHD) under a weak-field limit in Kerr spacetime, we present a set of preliminary results by numerically calculating the physical property of GRHD flows (e.g. kinematics and density distribution) in an effort to simulate redshifted absorption line spectra. Our model demonstrates that such GRHD accretion off the equatorial plane (i.e. $v \gsim 0.1c$ where $c$ is the speed of light in the vicinity of AGN closer than $\sim 100$ \sw radii) can manifest itself as UFIs in the form of redshifted absorption signature assuming the observed characteristics such as column density of $N_H \sim 10^{23}$ cm$^{-2}$ and ionization parameter of $\log (\xi \rm{[erg~cm~s^{-1}])} \sim 3$ as also seen in recent  multi-epoch {\it NuSTAR} observations among other data. 

\end{abstract}

\keywords{\uat{Active galactic nuclei}{16} --- \uat{Black hole physics}{159} --- \uat{High Energy astrophysics}{739} --- \uat{accretion}{14} --- \uat{Atomic spectroscopy}{2099} --- \uat{Plasma astrophysics}{1261}}

\section{Introduction} 

Accretion process plays a fundamental role in providing an efficient means to fuel central engines (aka. supermassive black holes) especially  in luminous active galactic nuclei (AGNs) where part of gravitational potential energy of accreting plasma is radiatively converted to thermal and nonthermal emission. On the other hand, the extensive spectroscopic observations in UV and X-ray band so far have clearly shown that a good fraction of such accreting materials can in fact manage to escape a black hole (BH) in the form of ionized outflows/winds via various physical mechanisms  \citep[e.g.][]{Crenshaw03,Gandhi22}. These powerful outflows, most likely originating from accretion disks of AGNs, have in fact been unambiguously identified
as blueshifted absorption lines from various ions depending on their line-of-sight (LoS) velocities
\citep[e.g.][]{Blustin05,Steenbrugge05, McKernan07}. In more recent years, it has become evident that many AGNs of diverse populations often exhibit ultra-fast outflows (UFOs), whose LoS velocity is near-relativistic (i.e. on the order of 10\% of the speed of light $c$),  in a highly variable fashion \citep[e.g.][]{Tombesi10,Pounds03, Reeves18, Chartas09, Nardini15, Parker17,Serafinelli19,Matzeu23,Gianolli24,Yamada24}. UFOs, being powerful outflows, are  hence believed to make a substantial impact on AGN feedback process \citep[e.g.][]{HopkinsElvis10} in BH-galaxy co-evolution.

Many of the observable signatures in general are  attributed to outflows (such as disk winds) and the medium located in narrow and broad line regions of AGNs. It is then tempting to search for some evidence of inflows or accretion itself at horizon scale. While observationally challenging, there have been a limited number of reports in literature using data with {\it Chandra}/gratings and {\it XMM-Newton} that imply the presence of infalling gas in the form of  redshifted absorption features in AGN X-ray spectra; e.g. Mrk~509 \citep{Dadina05}, E1821+643 \citep{YaqoobSerlemitsos05}, Mrk~335 \citep{Longinotti07}, PG1211+143 \citep{Reeves05, Pounds18, PoundsPage24}, and NGC~2617 \citep{Giustini17}. In these work, some authors employed  photoionization models  to constrain the physical property of ionized absorbers under thermal equilibrium, while others phenomenologically  adopted an inverted Gaussian function for spectral analysis. These studies thus allowed them to quantitatively characterize the likely inflows in these AGNs; i.e. near-relativistic inflow velocity (from redshift) of $v/c \sim 0.03-0.2$ and relatively high column density of  $N_H \gsim 10^{23}$ cm$^{-2}$ with somewhat high ionization state given by $\log (\xi \rm{[erg~cm~s^{-1}]}) \sim 3-4$. The inferred location of these fast inflows are often found to be in a close proximity to BH (i.e. on the order of  $r/R_g \sim 10-1,000$ where $R_g$ is gravitational radius). 
Nonetheless, these tentative detections have remained statistically elusive  due to limited significance (no more than $\sim 3\sigma-3.5\sigma$).

Besides AGNs, a small number of observations have implied similar spectral signatures of   redshifted absorption lines in 
transient X-ray binaries; e.g. MAXI~J1305-704 \citep{Miller14} and 4U~1916-053 \citep{Trueba20}, by spectral diagnostics based on a grid of photoionization models of multi-zone absorbers for multiple ions. 
Therefore, such fast inflows may be more common than currently thought possibly being a generic entity across mass scale. 

Interestingly, in a recent study of a heavily obscured Seyfert 2 AGN, ESP~39607 ($z=0.201$), \cite{Peca25} have reported multi-epoch {\it NuSTAR} detection of similar fast inflows at a higher significance ($\gsim 4\sigma$) as a result of photoionization modeling, revealing redshifted \fexxv\ He$\alpha$ absorption feature possibly blended with a minor contribution from \fexxvi\ Ly$\alpha$ line. The observed ultra-fast inflow (UFI) in their work is characterized by $v/c \sim 0.15-0.2$, $N_H \gsim 10^{23}$ cm$^{-2}$ and $\log \xi \sim 3$, probably launched from very small radius of $r/R_g \sim 20-100$.  

One of the natural interpretations of UFIs is associated with failed winds (aka. aborted jet) where gas initially   launched from a disk ends up falling back to the disk due to insufficient driving forces being unable to exceed local escape velocity. This phenomena has been demonstrated in a number of global numerical simulations in the context of hydrodynamic (HD) flows \citep[e.g.][]{Proga00,Proga04,Ghisellini04}.   
As another possibility, HD simulations have suggested that non-standard accretion flows may be produced by misaligned accretion disks that is subject to tearing instability, potentially causing the disk to break and allowing material to fall inward rapidly under chaotic accretion \citep[e.g.][]{Nixon12, GaspariSadowski17, Pounds18, Kobayashi18}.


Inspired by these theoretical speculations and likely UFI  detections, we propose in this paper an alternative scenario, perhaps in a more simplistic and natural context, where UFIs are identified, not as failed winds or chaotic accretion, but as relativistic accretion flows off the equatorial plane. In this framework, ionized accreting plasma  originating from an equatorial disk is   channeled externally by poloidal magnetic field  into low-to-mid-latitude region 
%
The lifted plasma, which is transonic, then continues to accrete along the field lines towards the event horizon of BH, which may be naturally perceived as UFIs.  
Such a BH magnetospheric structure governed primarily by poloidal magnetic field has been extensively investigated from a theoretical standpoint using global simulations with general relativistic (GR) magnetohydrodynamics (MHD) and force-free (vacuum) field, which remains among the outstanding problems to date  \citep[e.g.][]{TomimatsuTakahashi01,Hirose04,HawleyKrolik06,Fukumura04,Fukumura07,Punsly09,Ripperda22,Endo25}. 

Our present work here is hence motivated by both observational implications and theoretical suggestions in an effort to better provide a natural cause of the observed UFIs  and how they manifest themselves in X-ray spectra. In this paper, we consider for simplicity  stationary, axisymmetric GRHD accretion flows off the equatorial plane in Kerr geometry. By determining a number of GRHD accretion properties (namely, its kinematics and density distribution) for a given set of conserved flow quantities, we further compute some tangible spectral signatures. It is then  demonstrated that redshifted absorption features predicted in this model are in a broad agreement with the UFI observations. 

In \S 2, we briefly revisit the essential property of our non-equatorial GRHD accretion flow model. In \S 3, we present our results and spectral calculations based on a sample of fiducial GRHD solutions. We summarize and discuss our findings with implications in \S 4.


\begin{figure}[t]
\begin{center}
\includegraphics[trim=0in -0in 0in
0in,keepaspectratio=false,width=3.0in,angle=-0,clip=false]{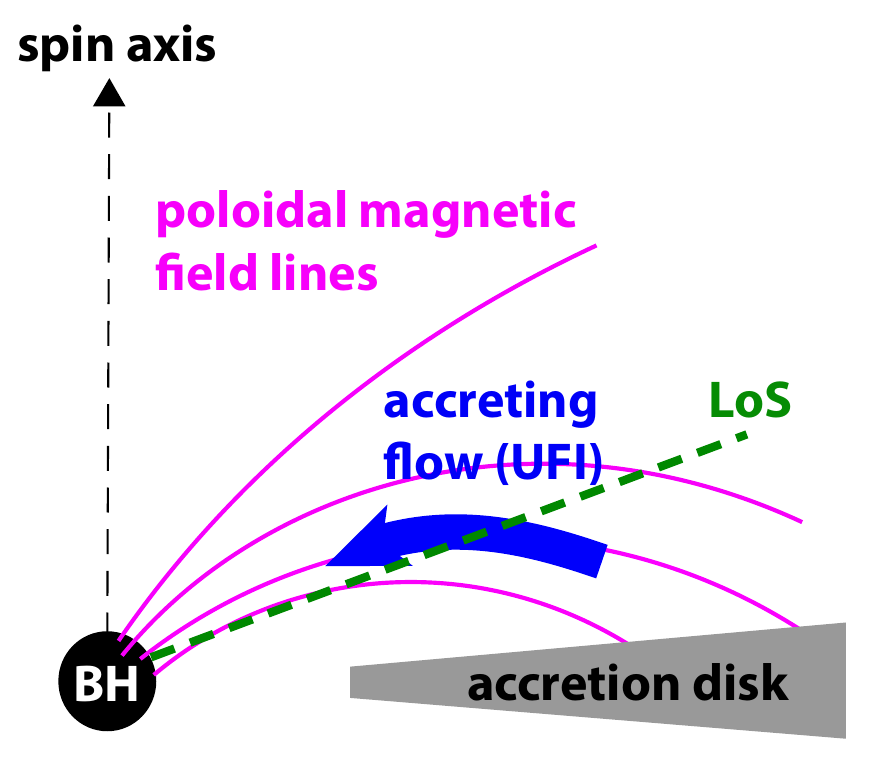}
\end{center}
\caption{Schematic picture of non-equatorial  accretion flows guided and channeled by the poloidal magnetic field (purple), which is simplified to be a conical geometry in our calculations so that $u^\theta=0$ for simplicity. Gas is provided from the equatorial disk surface at large distance. Green dashed line denotes line of sight (LoS) angle $\theta_{\rm obs}$. Note that toroidal motion of gas ($u^\phi \ne 0$) is suppressed in this poloidal projection. }
\label{fig:schematic}
\end{figure}

\section{Model Description} \label{sec:model}

Solving the geometry of BH magnetospheric structure  is an extremely difficult problem in a realistic fashion, and it has become computationally feasible only recently \citep[e.g.][]{Ripperda22}.
Since our primary objective is to explore a potential connection between an expected characteristics of relativistic inflows off the equator and the observed UFI spectra, we assume a conical inflow geometry  as an approximation  in our current methodology, which can be justified  especially at small radius where the field curvature becomes small enough.

\subsection{GRHD Accretion in Strong Gravity}

We consider steady state ($\partial/\partial t=0$), axisymmetric ($\partial/\partial \phi=0$) accreting flows in Kerr geometry. The spacetime metric is given by the conventional Boyer-Lindquist coordinates with the metric signature being $(-,+,+,+)$ such that the four-velocity normalization gives $\mathu \cdot \mathu=-1$ where $\mathu =(u^t,u^r,u^\theta,u^\phi)$.  Geometrized units are adopted for convenience  (i.e. $G=c=1$ where $G$ and $c$ are gravitational constant and the speed of light, respectively).  

\subsubsection{Characteristics of UFIs}

Providing that the presence of poloidal magnetic field threaded through the equatorial disk surface and the vicinity of the innermost stable circular orbit (ISCO),  we consider a conical UFI (i.e. $u^\theta=0$ and $u^r<0$) of an ideal Boltzmann gas  off the equatorial plane as an approximation of the actual streamline. Such an accreting gas spirals around BH spin axis ($u^\phi \ne 0$) because of angular momentum and accretes onto the BH of mass $M$. Poloidal projection of the inflows considered here is schematically illustrated in {\bf Figure~\ref{fig:schematic}}. 

As is often considered, a dynamical timescale of  accretion process is thought to be much shorter than that for the energy (or thermal) dissipation during the fluid accretion. Hence, the gas obeys the equation of state for an ideal gas, which can take the polytropic form as $P=K \rho^\gamma$ to link thermodynamic quantities such as  rest-mass density $\rho$ and thermal pressure $P$ of gas with adiabatic index $\gamma=4/3$. Here, $K$ has been used for a measure of entropy intimately related to locally measured gas temperature. Gas number density $n(r,\theta)$ is then expressed as  $n \equiv \rho/m_p$ where $m_p$ is the baryon mass in the flow.

Under these assumptions, specific total energy $E$ and axial angular momentum component $L$ of the gas are conserved along the fluid's trajectory such that $E \equiv -\mu u_t$ and $L \equiv \mu u_\phi$ where $\mu$ denotes relativistic enthalpy of gas and $u_{t,\phi}$ are  covariant components of $\mathu$ \citep[][]{Fukumura04,Fukumura07,Takahashi02,Takahashi08}.
%
 
%
%
With the local sound speed defined as $c_s \equiv (\partial P/\partial \epsilon)^{1/2}$ where $\epsilon \equiv \rho+\tilde{n}P$ is the gas total energy density, one finds that gas number density     scales as
\begin{eqnarray}
n(r,\theta) \propto \left[\frac{c_s(r,\theta)^2}{1-\tilde{n} c_s(r,\theta)^2} \right]^{1/(\gamma-1)} \ , \label{eq:density} 
\end{eqnarray}
where $\tilde{n} \equiv 1/(1-\gamma)=3$ in this work.
Total gas energy $E$ is  characterized as a combination of  sound speed (reflecting thermal component), radial four-velocity (reflecting kinetic component) and gravitational potential such that
\begin{eqnarray}
E = \left(\frac{1+u_r u^r}{-V_{\rm eff}(r,\theta;\ell)}\right)^{1/2}/(1-\tilde{n} c_s^2)  \  , \label{eq:energy}
\end{eqnarray}
where $u_r$ is a covariant radial four-velocity component and $V_{\rm eff}(r,\theta;\ell) \equiv g^{tt} - 2 \ell g^{t\phi} + \ell^2 g^{\phi\phi}$ is the effective potential  for inflows with specific angular momentum $\ell$ in Kerr metric of tensor components $g^{\alpha \beta}$ \citep[e.g.][]{Lu95,Fukumura07}. Note that the inflow angular momentum $L$ is then given as $L \equiv \ell E$. Here, prograde inflow is given by $\ell>0$ while retrograde inflow is expressed by $\ell<0$ with dimensionless BH spin parameter being $0 \le a \le 1$. In this formalism, mass inflow/accretion rate is conserved as well such that $\dot{\cal{M}} \propto n r^2 u^r$ assuming that the inflow is adiabatic throughout.  

By taking the radial derivative of equation~(\ref{eq:energy}),  the inflow radial velocity $u^r$ is numerically solved as a function of radius $r$ for a given conical angle (i.e. inflow inclination) $\theta$ \citep[e.g.][]{Chakrabarti89,Chakrabarti90,Fukumura04,Fukumura07}. The regularity condition demands that a physically valid inflow must be transonic; i.e. the gas makes a  transition from subsonic to supersonic state through a critical (sonic) point before crossing the horizon. Thus, the condition that $du^r/dr=\rm{finite}$ everywhere provides a sonic point at radius $r=r_c$ for physical UFIs. Be reminded that infalling gas is hence compressible,  adiabatic while preserving the initial mass accretion rate $\dot{\cal{M}}$ throughout the course of accretion in this formalism.

\subsubsection{Locally Flat Reference Frames}

With respect to an observer at infinity ($r \rightarrow \infty$), we  calculate physical gas velocity, i.e., three-velocity $\mathv=(v^r, v^\theta=0,v^\phi$), in two different locally-flat reference frames \citep[e.g.][]{Manmoto00}; (1) Locally nonrotating reference frame (LNRF) where a reference frame  corotates with the spacetime (i.e. rotating BH) subtracting the effect of frame-dragging. (2) Corotating reference frame (CRF) where a reference frame corotates with the gas around BH subtracting gas azimuthal motion. Primary components of physical three-velocity of UFI, $\mathv(r,\theta)$, is then given in each reference frame as
\begin{eqnarray}
v_{\rm LNRF}^\phi(r,\theta) &=& \frac{A}{\Sigma \Delta^{1/2}}  \left(\Omega-\omega \right) \sin \theta \  , \\
v_{\rm CRF}^r (r,\theta) &=& \left(\frac{u_r u^r}{1+u_r u^r} \right)^{1/2}   \  , 
\end{eqnarray}
where $\Delta \equiv r^2 + a^2 - 2 M r$, $\Sigma \equiv r^2 + a^2 \cos^2 \theta$, $A \equiv (r^2 + a^2)^2 - a^2 \Delta \sin^2 \theta$. 
In the above, gas angular velocity has been introduced as $\Omega \equiv u^\phi/u^t$   and the angular velocity of the dragging of the inertial frame is defined as $\omega \equiv -2Mar/A$.

\subsection{Absorption Spectrum of UFIs}

UFIs with the obtained kinematic property can imprint absorption features in the continuum X-ray spectrum as the UFIs are irradiated by  ionizing X-ray radiation (possibly  from a corona). 
While a global UFI density profile is explicitly given by equation~(\ref{eq:density}), its normalization needs to be determined. 
Hinted by a number of UFI observations\footnote[7]{As the current UFI sample size is not statistically sufficient, the values assumed in this paper may not accurately portrait the actual UFI population (but rather a preliminary attempt), which would be more robustly improved by future observations.} and photoionization modeling in literature, we adopt here $N_H = 10^{23}$ cm$^{-2}$ and $\log \xi = 3$ in a close proximity to BH at $r/R_g=50$ \citep[e.g.][]{Peca25}. 
These values are indeed consistent with those reported in the previous UFI studies mentioned above in \S 1. 
Then, the density normalization, $n(r=50R_g)$, is found to be 
\begin{eqnarray}
n(r=50R_g) = \frac{L_{\rm ion}}{r^2 \xi} \simeq  10^{11} ~ \rm{cm}^{-3}  \  , 
\end{eqnarray}
where $L_{\rm ion} \sim 10^{44}$ erg~s$^{-1}$ is ionizing X-ray luminosity in $1-10^3$ Ryd,  
following the case of ESP~39607 and other AGNs. 
Note that a different choice of  fiducial UFI location, $r$, thus would  yield a different density normalization in equation~(5). 
%
In parallel, volume filling factor of UFI is expressed by $b_{\rm vol}(r) = N_H/ [ n(r) r ]$. 

While hard to robustly identify in observations, we consider here \fexxvi\ Ly$\alpha$ 
line (6.97 keV) as a spectral signature of UFIs for simplicity adopting oscillator strength $f_{\rm ij}=0.7749$ and Einstein coefficient $A_{\rm ij}=5.03 \times 10^{14}$ s$^{-1}$. 
By employing Voigt profile $H(a,\nu)$ as a function of line photon frequency $\nu$ where $a \equiv A_{\rm ij} /(4 \pi \Delta v_D)$ and  $\Delta v_D$ denotes line broadening factor associated with turbulence and kinematic shear motion of gas,  photo-absorption cross section is determined by $\sigma_{\rm abs} = 0.01495 (f_{ij}/\Delta v_D) H(a,u)$. Typically, $\Delta v_D$ is $\sim 10\%$ of the bulk motion \citep[e.g.][]{F10}, as often found from UFO observations \citep[e.g.][]{Tombesi11}

UFI line optical depth  is then computed by $\tau(\nu) = \sigma_{\rm abs} N_{\rm ion}$ where $N_{\rm ion}$ is ionic column of UFIs (i.e. from \fexxvi\ in this case). Simulated line transmission due to UFIs is thus found by $e^{-\tau}$.
In the current formalism, $\tau(\nu)$ turns out to be independent of a choice of UFI density $n(r)$ for a fixed value of $N_H=10^{23}$ cm$^{-2}, \log \xi=3$ and $L_{\rm ion}=10^{44}$ erg~s$^{-1}$ because $\sigma_{\rm abs}$ is governed primarily by UFI velocity field once atomic property is provided (i.e. \fexxvi\ here). Accordingly, UFI line signature would be independent of $n(r)$. Likewise, the same argument applies for the effect of ionizing  luminosity $L_{\rm ion}$. Since $L_{\rm ion}$ is also disconnected to $\sigma_{\rm abs}$ here, UFI line feature does not depend on $L_{\rm ion}$ either. Instead, it is volume filling factor $b_{\rm vol}$ that would absorb the consequence. All in all, such a  decoupling is being allowed  due to the lack of radiative transfer process in the model. 


\begin{figure}[t]
\begin{center}
\includegraphics[trim=0in -0in 0in
0in,keepaspectratio=false,width=2.3in,angle=-0,clip=false]{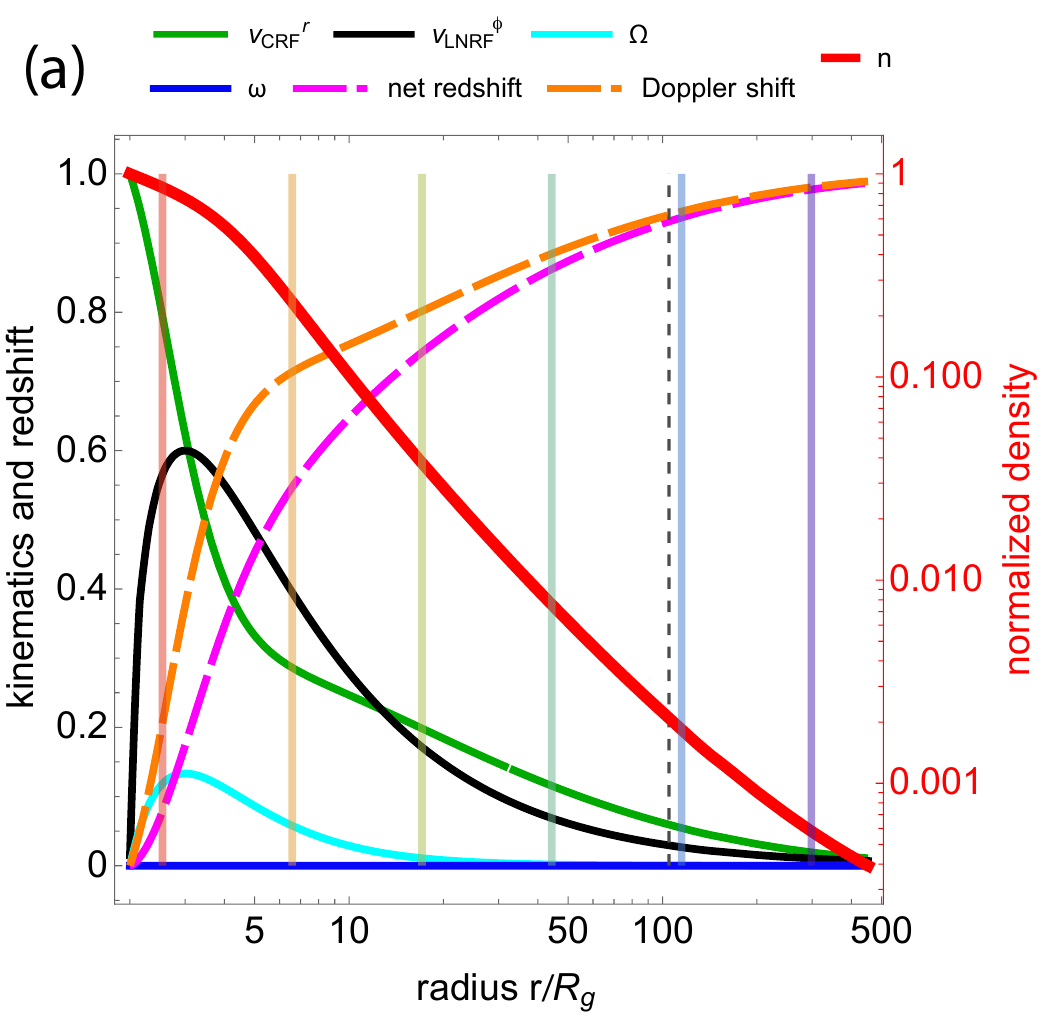}\includegraphics[trim=0in -0in 0in
0in,keepaspectratio=false,width=2.3in,angle=-0,clip=false]{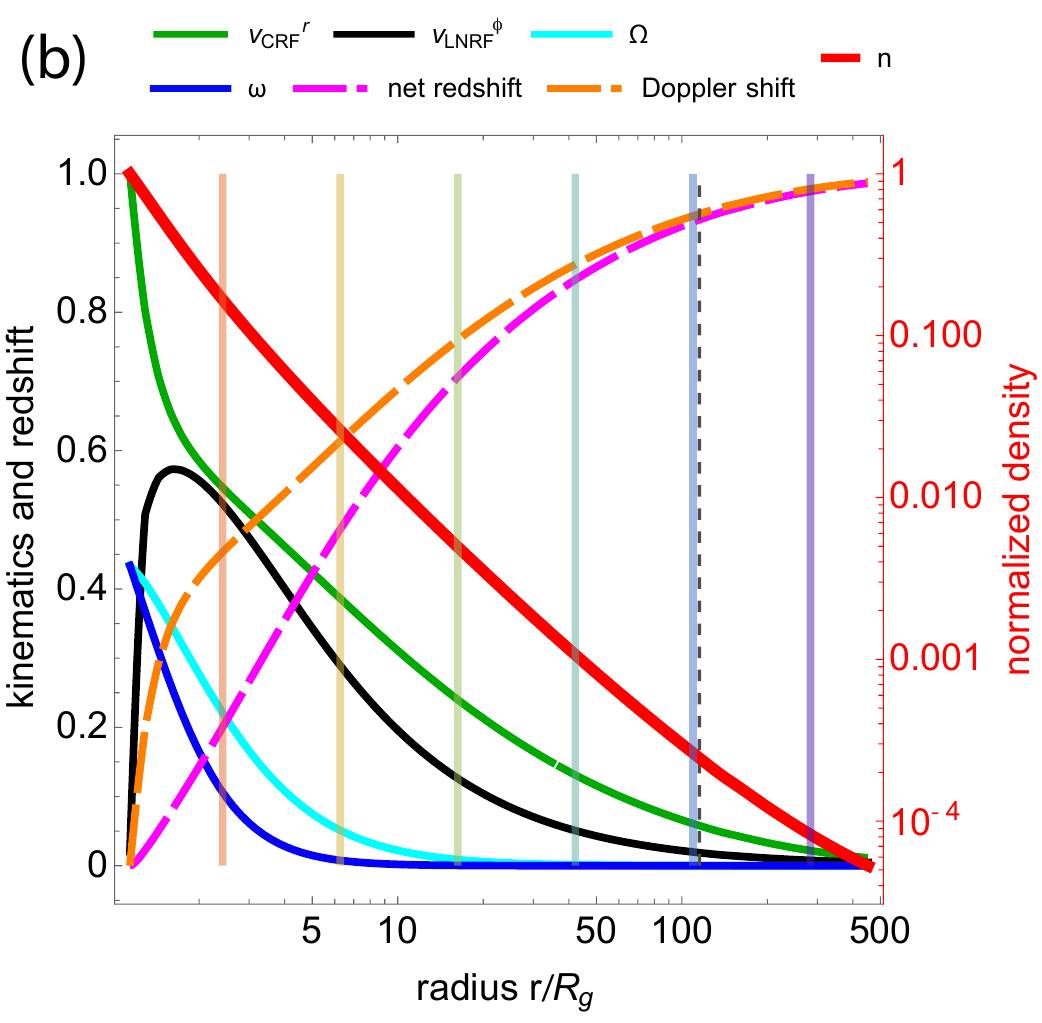}
\includegraphics[trim=0in -0in 0in
0in,keepaspectratio=false,width=2.3in,angle=-0,clip=false]{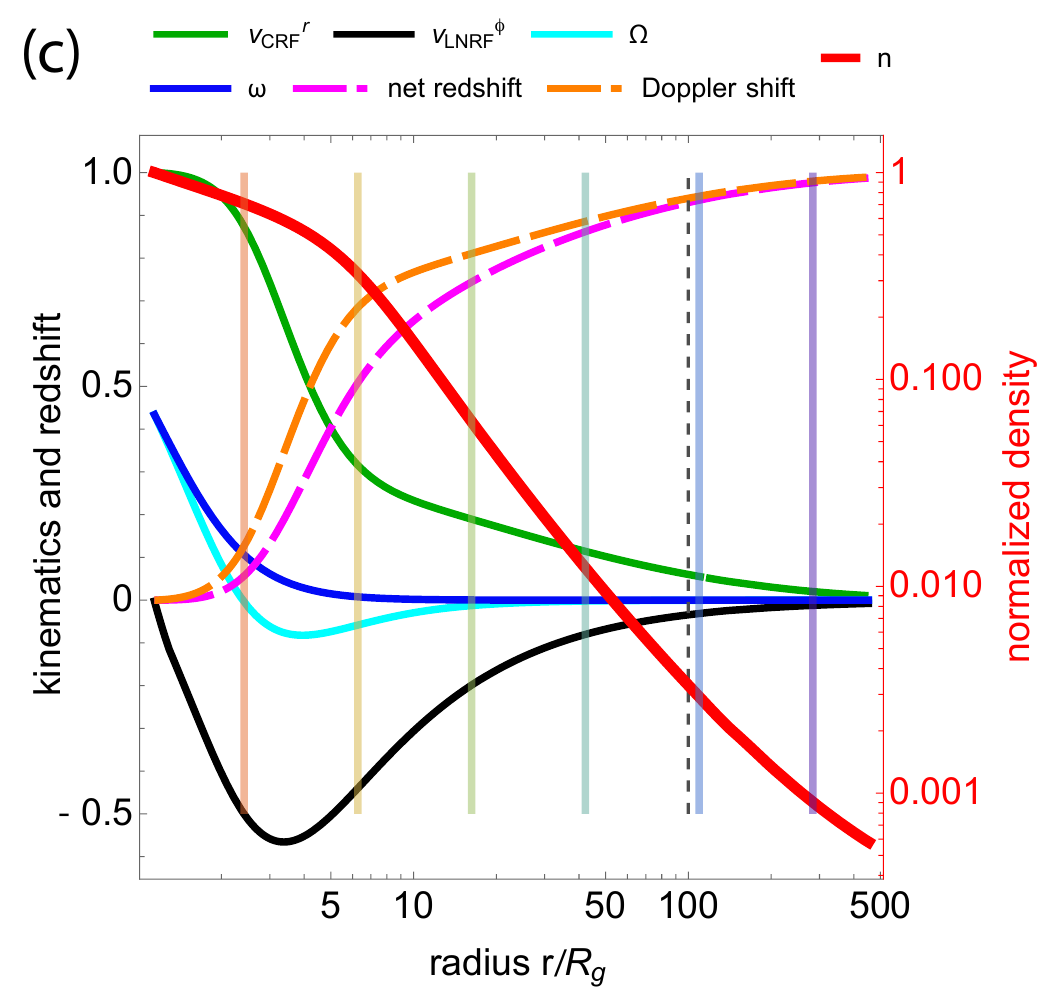}
\end{center}
\caption{Intrinsic kinematic solution of UFI as a function of distance $r$ for (a) a \sw BH ($a=0$) with $\ell=2.7$, (b) prograde UFI with $\ell=1.9$ and (c) retrograde UFI with $\ell=-3$ around a Kerr BH ($a=0.99$) with $\theta=60\deg$ showing $v_{\rm LNRF}^\phi/c$ (dark), $v_{\rm CRF}^r/c$ (green), inflow angular velocity $\Omega$ (cyan), and angular velocity of the inertial frame $\omega$ (i.e. frame-dragging; blue). Superimposed are the normalized number density $n(r)$ (thick red), net redshift $g_{\rm net}$ (dashed magenta) and classical Doppler shift $g_D$ (dashed orange) of UFIs. A sequence of color-coded vertical lines denote a series of sampled radii where we also compute spectra in {\bf Figure~\ref{fig:spec}}. Dotted vertical line indicates the sonic radius at $r=r_c \sim 100R_g$.}
\label{fig:accretion}
\end{figure}

\section{Results} \label{sec:results}

\subsection{Physical Condition of UFIs}

By specifying a set of fiducial quantities for accreting matter, we uniquely determine a global behavior  and the other physical property of UFIs. Following the earlier work on GRHD accretion, we  consider a conical GRHD inflow solution\footnote[8]{Similar GR(M)HD inflow solutions have been discussed in details  including the physical significance of parameters such as $E$ and $L$ \citep[e.g.][]{TomimatsuTakahashi01,Takahashi02}.} of $\theta=60\deg$ off the equatorial plane with $E=1.005$  \citep[e.g.][]{Fukumura04,Fukumura07}. 
{\bf Figure~\ref{fig:accretion}} shows an intrinsic kinematic solution of UFIs for (a) a \sw BH ($a=0$) with $\ell=2.7$, (b) prograde UFIs  with $\ell=1.9$ and (c) retrograde UFIs  with $\ell=-3$ around a Kerr BH ($a=0.99$). We show $v_{\rm LNRF}^\phi$ (dark), $v_{\rm CRF}^r$ (green), $\Omega$ (cyan) and $\omega$ (blue). We also present classical Doppler shift $g_D$ due to the radial inflow motion in CRF (dashed orange) and net redshift $g_{\rm net}$ (dashed magenta) which  includes GR gravitational shift as well. For reference,  normalized gas density $n$ (thick red) is also given.  In this scenario, accreting gas starts infalling at some distance as subsonic flow, gradually speeding up to become supersonic flow through a sonic point (vertical dashed line) at around $r_c/R_g \sim 100$ where $v_{\rm CRF}^r \sim c_S \sim 0.06c$ for three cases before entering the event horizon. 
A series of vertical lines in different color depicts different distances where we also calculate UFI absorption spectra (see \S 3.2).

It is seen that the radial motion of UFIs looks qualitatively similar in CRF (i.e. a locally flat reference frame) in all the cases. Rotational motion of UFIs, on the other hand, is different. Around a \sw BH in (a), the inflows  gradually spin up due to its intrinsic angular momentum $\ell$, but eventually ends up being fully radial motion  (i.e. $\Omega \sim v^\phi_{\rm LNRF} \rightarrow 0$  at the horizon) as expected. Consequently, we see $v^r_{\rm CRF} \rightarrow 1$ at the horizon \citep[e.g.][]{Abramowicz97,Manmoto00,Fukumura04}.

In the presence of frame-dragging due to BH spin in (b), prograde UFIs rotate faster than the local inertial frame everywhere  (i.e. $\Omega > \omega$), while eventually converging to each other at the horizon as required in LNRF (i.e. $v^\phi_{\rm LNRF} \rightarrow 0$ at $r \rightarrow R_g$; e.g. \citealt{Manmoto00,Lu98}). In the case of retrograde UFIs  in (c), the sense of gas rotation is opposite to that of Kerr BH (i.e. $a \ell<0$) clearly indicated by $\Omega<0$ and $v^\phi_{\rm LNRF}<0$ while $\omega>0$ initially. In the course of accretion, however, the frame-dragging becomes more effective near $r \gsim 2R_g$ (which is near the static limit) to the extent that the gas rotation gets reversed to corotate with Kerr BH (thus $\Omega$ changes its sign from $-$ to $+$) before entering the horizon where both gas and BH are corotating exactly at the same angular velocity (i.e. $\Omega=\omega$ yielding $v^\phi_{\rm LNRF}=0$ and hence $v^r_{\rm CRF}=1$ at the horizon)  as demanded by GR.     
%
%
Gas density  $n$ of UFIs is found to monotonically increase independent of BH spin because of increasing sound speed $c_s$ along accretion.

Classical Doppler shift $g_D$ depends sensitively on the radial UFI motion  $v^r_{\rm CRF}$ in CRF. Infalling gas is also subject to GR gravitational redshift in Kerr spacetime independent of the kinematic motion. While GR redshift is negligible at large distance away from BH, it becomes more dominant over the Doppler effect at small radius because GR redshift approaches zero at the horizon. For this reason, the net redshift $g_{\rm net}$ becomes more significant  towards the horizon regardless of the intrinsic motion of UFIs.    

We have confirmed that  the effect of conical angle $\theta$ on inflow property is only minimum not drastically changing the UFI solutions. This is mainly because 
gas kinematics is primarily governed by energy $E$, not angle $\theta$. 
%
Although nontrivial to make a precise quantitative estimate of the effect of field line geometry in the current GRHD framework, we naively speculate that a different field structure with distinct curvature would not make a dramatic difference either in UFI characteristics as long as gas energy is kept more or less the same (the field lines are almost radial  in a close proximity to AGN). We also point out, unlike outflows/winds, that strong gravity plays a role more effectively near AGN as considered in this work, alleviating the effects by other external factors like magnetic field and gas/ram pressure unless the inflow energy or the field strength is substantially different.

\begin{figure}[t]
\begin{center}
\includegraphics[trim=0in -0in 0in
0in,keepaspectratio=false,width=2.3in,angle=-0,clip=false]{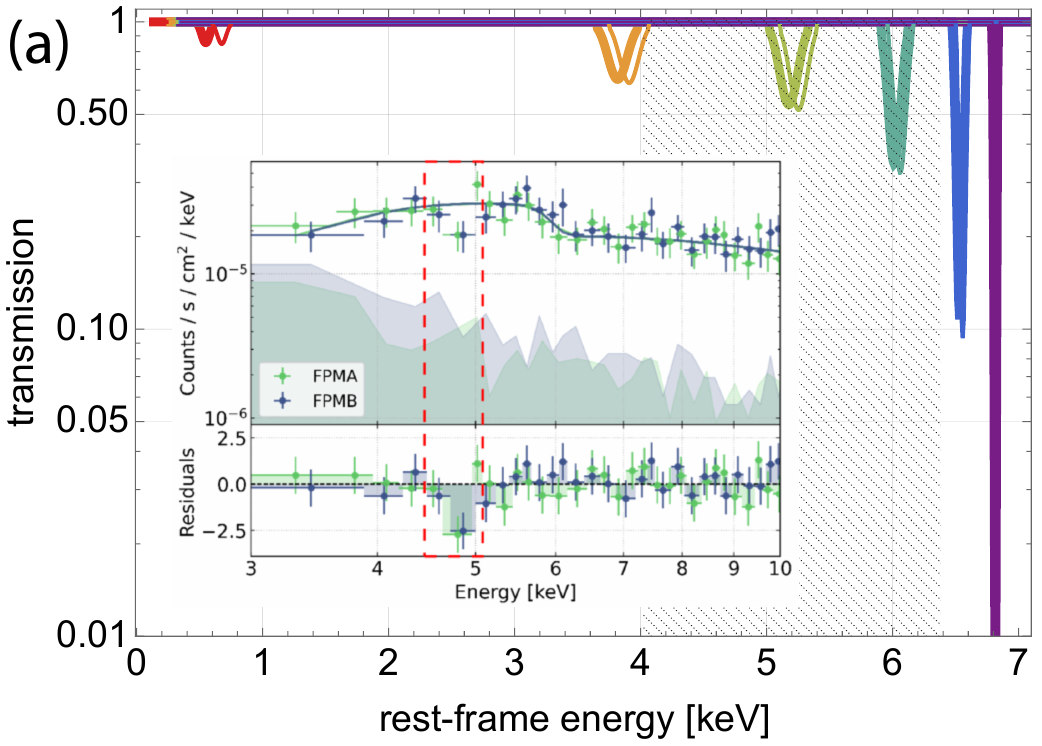}\includegraphics[trim=0in -0in 0in
0in,keepaspectratio=false,width=2.3in,angle=-0,clip=false]{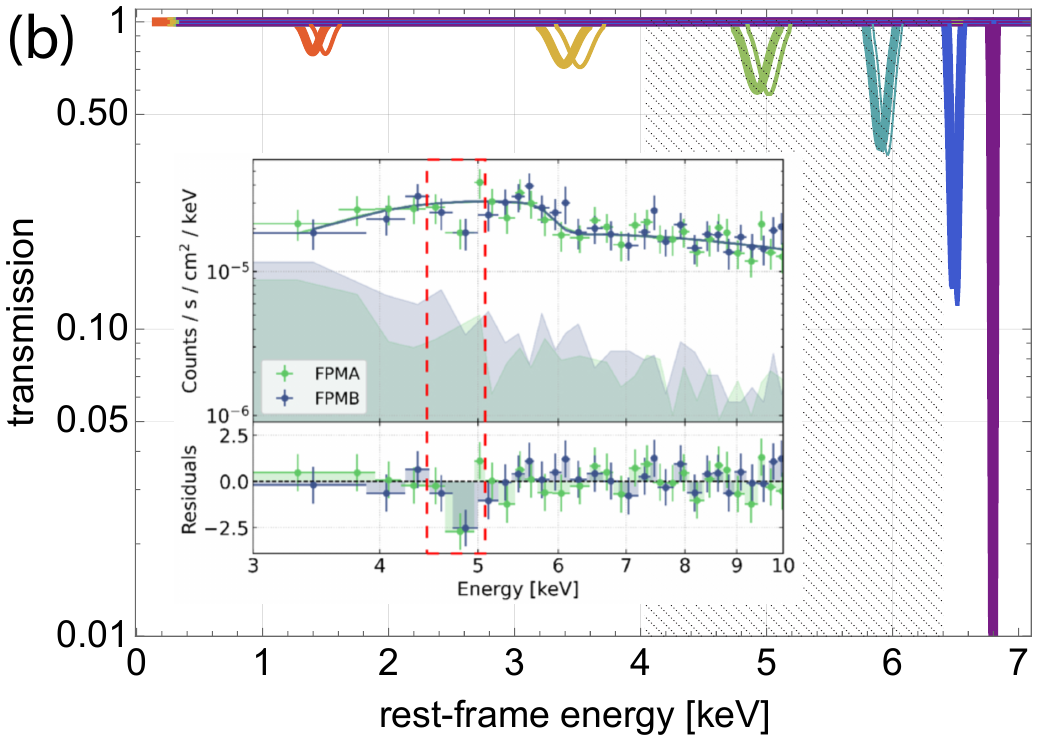}
\includegraphics[trim=0in -0in 0in
0in,keepaspectratio=false,width=2.3in,angle=-0,clip=false]{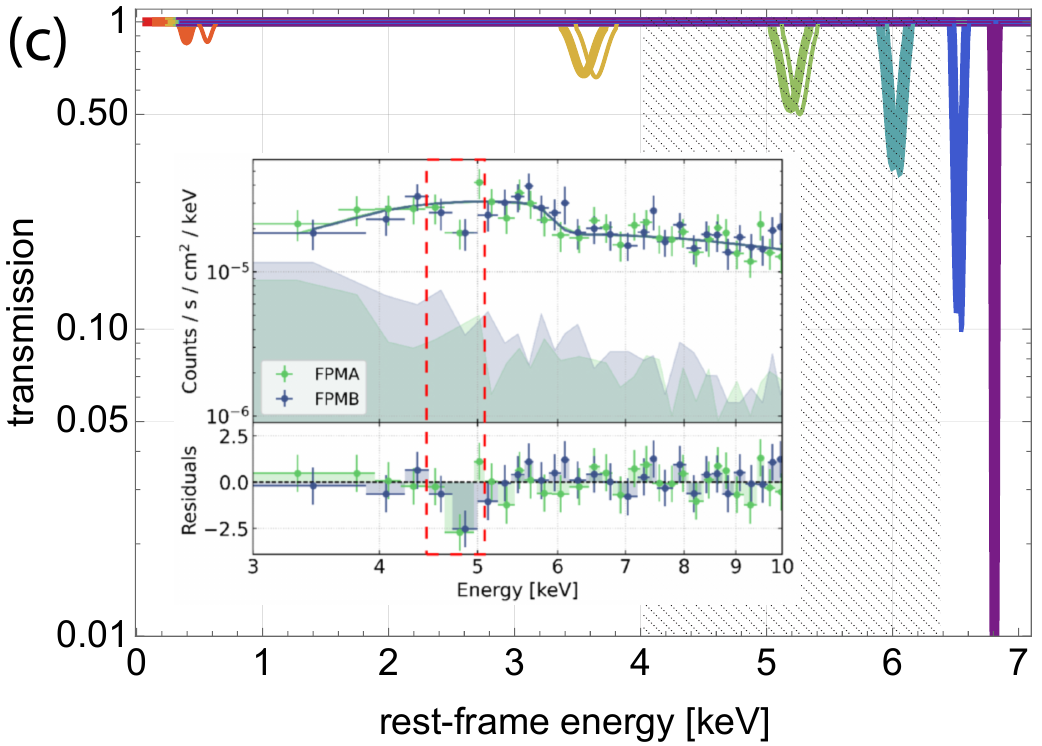}
\end{center}
\caption{Calculated \fexxvi\ Ly$\alpha$ absorption line  spectra (transmission) for (a) a \sw BH ($a=0$), (b) prograde UFI  and (c) retrograde UFI around a Kerr BH ($a=0.99$)  with $\theta=60\deg$ at various radii corresponding to those  (a sequence of color-coded vertical lines) shown  in {\bf Figure~\ref{fig:accretion}}. We show $\theta_{\rm obs}=60\deg$ (thick) and $40\deg$ (thin) for comparison.  Shaded region depicts the energy band of the {\bf detected} UFIs derived from multiple AGN observations.  Inset is {\it NuSTAR} spectrum of ESP~39607 exhibiting a potential UFI feature (red rectangle) for reference \citep[adopted from ][]{Peca25}. }
\label{fig:spec}
\end{figure}

\subsection{Absorption Spectrum of UFI}

Given the calculated kinematics of UFI with the corresponding net redshift factor $g_{\rm net}$, a sequence of absorption lines of \fexxvi\ Ly$\alpha$ from   UFI is simulated as a function of distance in {\bf Figure~\ref{fig:spec}}. Each absorption feature is produced by UFI  at different radius corresponding to color-coded vertical line shown in {\bf Figure~\ref{fig:accretion}}. 


For simplicity in the present toy model\footnote[9]{In a realistic photoionization modeling, line width and depth of  absorption feature would be sensitive to $\xi$.}, we perform spectral calculations assuming that UFI is globally characterized by a constant column density of $N_H = 10^{23}$ cm$^{-2}$  (corresponding to $n \simeq 10^{11}$ cm$^{-3}$ at $r /R_g=50$) and $\log \xi =3$ at every radius, being motivated by a series of observations as addressed in \S 2.2. Our calculations demonstrate that part of accreting UFI located at different radius can absorb X-ray  (originating from coronae) being identified as UFI in this framework.   
Due to the uncertainty in relative angle between our LoS and the inflow geometry, we consider two LoS inclinations; $\theta_{\rm obs}=60\deg$ (same as the UFI's conical angle; thick) and $40\deg$ (thin) in {\bf Figure~\ref{fig:spec}} assuming a UFI conical angle of $\theta=60\deg$.   Shaded region depicts the energy band of the detected UFIs derived from observations of multiple AGNs.  Again, the color sequence of each spectrum  corresponds to that of inflow radius (a series of vertical lines) shown in {\bf Figure~\ref{fig:accretion}}. 


With the low initial inflow velocity at large distance in all cases, the absorption line is very narrow  due to smaller turbulence $\Delta v_D$ with little redshift as expected. With increasing velocity towards BH, line energy shifts to lower energy due to more redshift (both classical Doppler and gravitational). At the same time, the line becomes broader due to increasing  turbulence $\Delta v_D$. It is seen that the centroid line energy of \fexxvi\ Ly$\alpha$ becomes extremely redshifted (e.g. $E \lsim 6$ keV) when produced from UFI at  small distance (e.g. $r/R_g \lsim50$) due mainly to GR gravitational redshift in addition to classical Doppler redshift. 
Note that a more realistic treatment of the evolution of ionization state of UFI with distance is beyond the scope of this paper.
Closer to the horizon, GR gravitational redshift starts dominating over classical Doppler shift, which is slightly different in (a)-(c) depending on radial motion of UFI (given by $v^r_{\rm CRF}$) due to different angular momentum $\ell$. Hence, the exact line shift is slightly different for (a)-(c) at a given radius. Frame-dragging  plays little role in drastically reshaping absorption feature in all cases because radial motion dominates over toroidal component.   
Near the horizon where $v^r_{\rm CRF} \rightarrow 1$, turbulent motion $\Delta v_D$ becomes so large  that the photo-absorption cross section $\sigma_{\rm abs}$ becomes very small in a way that the line spectrum turns to be weaker/narrower. The line becomes even more redshifted but much weaker at $E \lsim 4$ keV for $r/R_g \lsim 10$ perhaps at an observationally unnoticeable level. 
As clearly seen, X-ray data from multiple AGNs  (indicated by shaded region), including ESP~39607 (inset), disfavors certain range of radial location of UFIs; i.e. $r/R_g \lsim 10$ and $r/R_g \gsim 80$ in this preliminary modeling purely based on predicted redshift of line energy.  

In comparison with those predicted for  LoS of $\theta_{\rm obs}=60\deg$  (thick), the line shift is found to be slightly smaller due to a smaller  velocity component projected into LoS of $\theta_{\rm obs}=40\deg$ (thin) in (a)-(c), although the difference in redshift is small at large distance due to smaller $g_{\rm net}$ as expected. With decreasing radius, the difference in redshift tends to be enhanced. 


\begin{figure}[t]
\begin{center}
\includegraphics[trim=0in -0in 0in
0in,keepaspectratio=false,width=3.6in,angle=-0,clip=false]{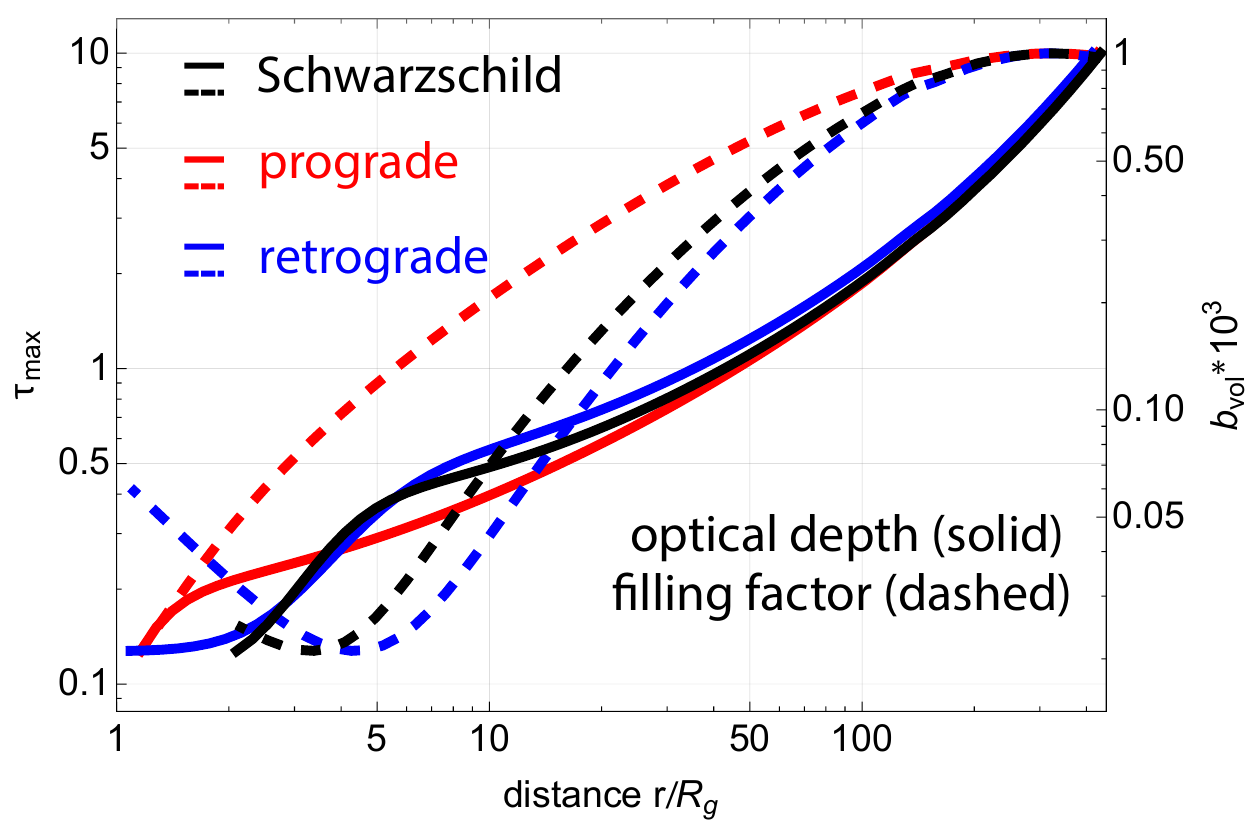}
\end{center}
\caption{Calculated  maximum line optical depth $\tau_{\rm max}$ (solid) and volume filling factor $b_{\rm vol}$ (dashed) of UFIs  as a function of radius $r$ for a \sw BH (black), prograde UFI (red) and retrograde UFI (blue), corresponding to {\bf Figures~\ref{fig:accretion}-\ref{fig:spec}}. Here, we assume a constant column $N_H=10^{23}$ cm$^{-2}$ with $\log \xi=3$ in the course of accretion for simplicity. }
\label{fig:soln}
\end{figure}

To complement the nature of UFI that is considered in this work, we compute the maximum line optical depth $\tau_{\rm max}$. In addition, because we assume a constant column and ionization parameter of UFI in our spectral calculations, volume filling factor $b_{\rm vol}$ is required to vary with radius. In  {\bf Figure~\ref{fig:soln}}  $\tau_{\rm max}$ (solid) and $b_{\rm vol}$ (dashed) are shown as a function of radius $r$  for a \sw BH (dark), a prograde UFI (red) and a retrograde UFI (blue).  
Again, assuming a constant column $N_H=10^{23}$ cm$^{-2}$ with $\log \xi=3$ in the course of accretion, we see that the gas becomes more clumpy or sparse (i.e. $b_{\rm vol} \sim 2 \times 10^{-5} - 10^{-3}$) with decreasing radius. This is mainly because gas density increases faster than decrease in radius. 
UFI becomes more optically thin with decreasing radius  even though we assume a constant column because photo-absorption cross section $\sigma_{\rm abs}$ becomes smaller due to increasing kinematic velocity and its turbulence. We note, however, that these quantitative estimates are highly parameter dependent in our calculations, while its qualitative trend is roughly generic to the model setup here. This property would also be qualitatively distinct if we relax the present assumption of a constant column and ionization in a more realistic situation under photoionization balance. 


\section{Summary \& Discussion} \label{sec:summary}

In this work, we consider  a steady-state GRHD conical inflows off the equatorial plane in Kerr metric under axisymmetry as a plausible candidate for UFIs hinted from a number of X-ray observations in AGNs (and some BH XRBs). By utilizing fiducial GRHD solutions (i.e. kinematics and density profiles with distance) for various BH spin parameters, absorption lines originating from UFI are computed assuming that the feature is attributed  to \fexxvi\ Ly$\alpha$  with constant column density of $N_H = 10^{23}$ cm$^{-2}$ and  high ionization parameter of $\log \xi =3$ in the course of accretion. A sequence of redshifted absorption features  is calculated as a function of distance to demonstrate the characteristic line spectrum in the current GRHD formalism under a weak magnetic field approximation. In this preliminary effort, we show that the rest-frame absorption feature can be widely redshifted depending on the location of inflows (i.e. UFIs), allowing a broad range of predicted energy, $E \sim 4-6.5$ keV, over the radial extent of $r/R_g \sim 10-80$ where UFI accretes at $v/c \sim 0.1-0.3$ in the current framework. This is in a broad agreement with X-ray observations of a diverse AGN population exhibiting UFIs \citep[e.g.][]{Dadina05,YaqoobSerlemitsos05,Reeves05, Longinotti07, Giustini17, Pounds18, PoundsPage24, Peca25}.

In this GRHD framework, we obtain a unique transonic solution with a unique sonic point for a given set of $(E, \ell, \theta; a)$ as we consider a complete inflow that is released from accretion at some distance (i.e. $r/R_g \sim 500$) falling into the event horizon without developing shocks along the way.  Such shock-free inflow solutions are thus restricted to a certain set of ($E,\ell$).
Therefore, depending on the combination of these inflow parameter vales, one will obtain a uniquely distinct solution with different kinematic and thermal property. From a theoretical standpoint, we have very little handle to constrain which solution would be more favored among others, thus allowing some degree of degeneracy. To this end, UFI observations become very informative because the observed redshifted absorption features can be used as a tangible proxy to determine which solution of those would be observationally more viable. Although this is  beyond the scope of our present work, we have shown in this  work a preliminary implication and it will certainly be better explored in a subsequent work. 

It has been postulated so far that the observed UFIs could be a signature of the so called failed winds or aborted jet that is initially  launched from the disk surface, resulting in falling back inwards due to the lack of sufficient outwards momentum and/or driving force at some distance \citep[e.g.][]{Proga00,Proga04,Ghisellini04,Nixon12}. In this viewpoint,  winds are partially, if not fully, transformed to inflows (thus UFIs) causing the observed redshifted absorption. For example,  a previous observation of ESP~39607 with {\it Suzaku} in 2010,  which took place $\sim 13$ years prior to  the first {\it NuSTAR} detection, 
 has indicated no sign of UFIs \citep[e.g.][]{Ricci17,Peca25}, perhaps implying a transient nature expected from such failed winds and/or chaotic accretion \citep[e.g.][]{Nixon12, GaspariSadowski17, Pounds18, Kobayashi18}.
The observed UFIs in the past observations are near-relativistic (i.e. $v/c \sim 0.1-0.2$). It is unclear how and if the outflows launched out to, say, $r/R_g \sim 100$ would in fact reverse its direction and further reach such high velocities coming back inwards. While very tempting, a detailed nature and physical property of the observed UFIs are therefore not entirely accounted for by the concept of failed winds. 
Multi-epoch {\it NuSTAR} observations of ESP~39607 in 2023 and 2024 (more than one year apart), on the other hand, have indicated a persistent presence of UFIs with physically similar properties like column density and line redshift \citep[][]{Peca25}, perhaps suggesting that UFIs may be dynamically sustained over a long timescale (e.g. months-years).  
A recent X-ray study of Mrk~3 has indeed indicated a persistent presence (over 11 years) of  ionized inflows of materials  at sub-pc scale between a putative torus and the outer accretion disk \citep{Shi25}. UFIs considered in this paper, on the other hand, could be inflows connecting the inner disk and BH/ionizing source  as conceived in {\bf Figure~\ref{fig:schematic}}.
To robustly explore a variable nature of UFIs, more multi-epoch observations will be needed.

In this work, we propose an alternative explanation for a physical identity and the property of the observed  UFIs in terms of conical accretion flows off the equatorial plane  where plasma gas is guided and channeled by a global poloidal magnetic field as illustrated in {\bf Figure~\ref{fig:schematic}}. 
In general,  MHD accretion can be hydro-dominated or magneto-dominated \citep[e.g.][]{Takahashi02,Takahashi08}. However, the MHD case is generally very complicated and, therefore, the main motivation of our current paper, as a preliminary effort, is to investigate the HD limit, which should be valid in the case of small magnetization. 
Hence,  we employ a model that should apply to the hydro-dominated inflows where the magnetic field does not make a significant contribution to the properties of inflows under a weak-field limit \citep[e.g.][]{Fukumura04}. 
Under such a condition, perturbing magnetic forces to the gas being responsible for major acceleration in $\theta$-direction is assumed to be negligible, keeping the inflow at constant in a conical flow (i.e. $\theta=$const). This approximation should apply to flows with small magnetization parameter perhaps on the order of $\sigma_M \sim 10^{-4}-0.1$ (where $\sigma_M$ is the ratio of Poynting flux to fluid mass flux) as often discussed in literature based on numerical calculations \citep[e.g.][]{Datta24, Fukumura04, Jacquemin-Ide19}. 
In such a case, HD should primarily control the nature of inflows (i.e. UFIs). In a subsequent paper, we plan to extend the model further to handle (sufficiently) magnetized UFIs  within a more rigorous GRMHD framework \citep[e.g.][]{Fukumura07,Takahashi02,Takahashi08}.

One caveat in our model is the lack of considering radiative transfer process. For a given ionizing X-rays (i.e. ionizing spectral energy distribution from a disk and corona), one can solve for an ionization balance under thermal equilibrium of illuminated UFIs in an effort to calculate expected Fe column density $N_H$, ionization parameter $\xi$ and gas temperature $T$, for example. To illustrate a feasible mechanism  to produce UFIs in this work, we ignore this process by assuming a constant $N_H=10^{23}$ cm$^{-2}$ and $\log \xi =3$ as a preliminary study. Given the uncertainty in the morphology and physical identity of UFIs, we also hold them constant throughout accretion. 

By carrying out radiative transfer calculations, it is conceivable that there would be an optimal range of radius where \fexxvi\ Ly$\alpha$  can be predominantly produced with a proper ionization front. For example,  little/weaker absorption would be expected at both large and small distances due to density and ionization equilibrium. Hence, a strong and prominent absorption line could be most likely imprinted somewhere in the middle in distance. 
Indeed, as the gas gets closer to the X-ray source, it naturally tends to be more  ionized unless density varies accordingly. As a possible scenario, the gas could be initially at lower ionization state (e.g. \fexx-\fexxiv) at larger distance. In that case, the line would be identified  as \fexxv-\fexxvi\ later at smaller radius when the gas gets closer to the X-ray source as ionization state is related to distance. 
Consequently, it is plausible to predominantly observe high-ionization  lines with faster velocities ($v/c \sim 0.1-0.2$) from smaller distance and low-ionization lines  with lower velocities ($v/c < 0.1$) from larger distance, for example.
Once $N_H$ and $\xi$ are physically allowed to vary in response to ionizing flux in radiative transfer calculations,  synthetic absorption line due to UFIs would be uniquely obtained for a given inflow, depending on how $N_H$ and $\xi$ change with distance $r$. This information is then used to more realistically simulate UFI spectrum. Again, the present methodology  merely provides a preliminary characterization of UFIs without incorporating radiative transfer process, and it will be improved significantly once the current assumption is relaxed.    
%


Regarding another characteristics of the predicted UFIs, the estimated small $b_{\rm vol}$ may physically imply clumpy nature as recently observed in X-ray UFOs from a nearby quasar, PDS~456, with XRISM/Resolve \citep[][]{XRISM25} where the authors have modeled the property of clumpy UFOs with a smooth HD framework.
We would similarly argue here that kinematics and thermal nature of clumpy UFIs may well still obey the fundamental nature of continuous GRHD inflows. It is plausible that clumpy gas originates from a smooth, continuous inflow while segments of such inflows get gradually and episodically peeled off as the gas accretes \citep[e.g.][]{Waters21,Mehdipour25}. 
Similar disintegration can be expected for clumpy UFIs.

In conjunction with other type of observed outflows (e.g. warm absorbers and UFOs) in AGNs, one might beg a question as to why we only see a handful of UFI detections to date, while UFOs are apparently far more ubiquitous ($\sim 40\%$ of  AGNs, see \citealt{Tombesi10, Gofford13,Matzeu23,Gianolli24,Yamada24}). UFOs are generally seen beyond the horizon scale (e.g. $r/R_g \gg 10$), while UFIs are even closer to AGNs. One can  argue that  it is generally harder to find/detect spectroscopic signatures  from smaller radius due to many external factors; e.g. diluted flux due to gravitational potential, a rapid variability to hide an underlying coherence in spectral lines, and so on. 
Hence, it is conceivable that UFOs and UFIs may not have an equity in detection rate/probability primarily because of the location. 
Nonetheless, 
the far smaller detection rate of UFIs when compared to that of UFOs poses a fundamental question;
Are UFIs physically real and if so, what is the estimation of expected UFIs/UFOs ratio?
Although the present model in this work with GRHD formalism cannot directly answer this question, an apparent  inequality between UFO and UFI detections in literature so far might be due to observational limitations if not due to a truly intrinsic difference. 

Lastly, dynamical stability of magnetic field at the horizon scale (e.g. near the ISCO) can be  short given the magnetic turbulence. For example, magnetized accreting plasma in global MHD simulations near the inner edge (i.e. ISCO) is found to exhibit accretion rate history with rapid variability  with turbulent nature swinging from ``low" flux state to ``high" flux state, perhaps characterized by a doubling time of $t \sim 500 R_g/c \sim 25$ ksec for a $10^7\Msun$ AGN \citep[e.g.][]{HawleyKrolik02,Hirose04}. This potentially implies a rapid duty cycle of MHD-driven UFIs in the innermost accretion region if the poloidal  field is spatially connected to small disk radius, for which multi-epoch observations could confirm the appearance and disappearance of UFIs. 
That is, we would expect a relatively short-term variable UFIs in terms of line width (due to turbulent kinematic nature) and depth (due to variable density) as well. Such a rapid variability predicted in these numerical simulations is in fact consistent with the  seemingly episodic UFI detections in the current limited observations.
On the other hand, if the field originates from a distant part of the disk ($r/R_g \gg 100$), then it is conceivable that expected UFIs may be magnetically sustained for a longer timescale. In this case, spectral signature of UFIs should be persistent over years. Given the lack of coordinated observations of UFIs to date, it would be insightful to obtain more data over both short and long timescales. 
Future observations of a detailed spectroscopy, especially with XRISM, will be thus crucial to obtain more direct evidence of UFIs along with their characteristic property.


\begin{acknowledgments}

We thank an anonymous referee for illuminating questions and comments to improve the quality of this manuscript. 
KF is grateful to Masaaki Takahashi for an illuminating discussion GRMHD accretion in the context of  BH magnetosphere. RS acknowledges funding from the CAS-ANID grant number CAS220016.


\end{acknowledgments}

\begin{contribution}

KF designed and initiated the project being motivated by {\it NuSTAR} data analyzed by AP and RS while also inspired by {\it BeppoSAX} and {\it XMM-Newton} data studied by MD.  All authors contributed to explanation and discussion, followed by preparing the manuscript.


\end{contribution}

%
\facilities{NuSTAR, XMM-Newton, Suzaku, Chandra }

\software{}


%
%
%


\bibliography{UFI}{}

@ARTICLE{Mehdipour25,
       author = {{Mehdipour}, Missagh and {Kaastra}, Jelle S. and {Eckart}, Megan E. and {Gu}, Liyi and {Ballhausen}, Ralf and {Behar}, Ehud and {Diez}, Camille M. and {Fukumura}, Keigo and {Guainazzi}, Matteo and {Hagino}, Kouichi and {Kallman}, Timothy R. and {Kara}, Erin and {Li}, Chen and {Miller}, Jon M. and {Mizumoto}, Misaki and {Noda}, Hirofumi and {Ogawa}, Shoji and {Panagiotou}, Christos and {Tanimoto}, Atsushi and {Zhao}, Keqin},
        title = "{Delving into the depths of NGC 3783 with XRISM: I. Kinematic and ionization structure of the highly ionized outflows}",
      journal = {\aap},
         year = 2025,
        month = jul,
       volume = {699},
          eid = {A228},
        pages = {A228},
          doi = {10.1051/0004-6361/202555623},
archivePrefix = {arXiv},
       eprint = {2506.09395},
 primaryClass = {astro-ph.HE},
       adsurl = {https://ui.adsabs.harvard.edu/abs/2025A&A...699A.228M}
}

@ARTICLE{Waters21,
       author = {{Waters}, Tim and {Proga}, Daniel and {Dannen}, Randall},
        title = "{Multiphase AGN Winds from X-Ray-irradiated Disk Atmospheres}",
      journal = {\apj},
         year = 2021,
        month = jun,
       volume = {914},
       number = {1},
          eid = {62},
        pages = {62},
          doi = {10.3847/1538-4357/abfbe6},
archivePrefix = {arXiv},
       eprint = {2101.09273},
 primaryClass = {astro-ph.GA},
       adsurl = {https://ui.adsabs.harvard.edu/abs/2021ApJ...914...62W}
}

@ARTICLE{Jacquemin-Ide19,
       author = {{Jacquemin-Ide}, J. and {Ferreira}, J. and {Lesur}, G.},
        title = "{Magnetically driven jets and winds from weakly magnetized accretion discs}",
      journal = {\mnras},
         year = 2019,
        month = dec,
       volume = {490},
       number = {3},
        pages = {3112-3133},
          doi = {10.1093/mnras/stz2749},
archivePrefix = {arXiv},
       eprint = {1909.12258},
 primaryClass = {astro-ph.HE},
       adsurl = {https://ui.adsabs.harvard.edu/abs/2019MNRAS.490.3112J}
}

@ARTICLE{Datta24,
       author = {{Datta}, Sudeb Ranjan and {Chakravorty}, Susmita and {Ferreira}, Jonathan and {Petrucci}, Pierre-Olivier and {Kallman}, Timothy R. and {Jacquemin-Ide}, Jonatan and {Zimniak}, Nathan and {Wilms}, Joern and {Bianchi}, Stefano and {Parra}, Maxime and {Clavel}, Ma{\"\i}ca},
        title = "{Impact of disc magnetisation on MHD disc wind signature}",
      journal = {\aap},
         year = 2024,
        month = jul,
       volume = {687},
          eid = {A2},
        pages = {A2},
          doi = {10.1051/0004-6361/202349129},
archivePrefix = {arXiv},
       eprint = {2403.13077},
 primaryClass = {astro-ph.HE},
       adsurl = {https://ui.adsabs.harvard.edu/abs/2024A&A...687A...2D}
}

@ARTICLE{XRISM25,
       author = {{XRISM Collaboration} and {Audard}, Marc and {Awaki}, Hisamitsu and {Ballhausen}, Ralf and {Bamba}, Aya and {Behar}, Ehud and {Boissay-Malaquin}, Rozenn and {Brenneman}, Laura and {Brown}, Gregory V. and {Corrales}, Lia and {Costantini}, Elisa and {Cumbee}, Renata and {Trigo}, Mar{\'\i}a D{\'\i}az and {Done}, Chris and {Dotani}, Tadayasu and {Ebisawa}, Ken and {Eckart}, Megan and {Eckert}, Dominique and {Enoto}, Teruaki and {Eguchi}, Satoshi and {Ezoe}, Yuichiro and {Foster}, Adam and {Fujimoto}, Ryuichi and {Fujita}, Yutaka and {Fukazawa}, Yasushi and {Fukushima}, Kotaro and {Furuzawa}, Akihiro and {Gallo}, Luigi and {Garc{\'\i}a}, Javier A. and {Gu}, Liyi and {Guainazzi}, Matteo and {Hagino}, Kouichi and {Hamaguchi}, Kenji and {Hatsukade}, Isamu and {Hayashi}, Katsuhiro and {Hayashi}, Takayuki and {Hell}, Natalie and {Hodges-Kluck}, Edmund and {Hornschemeier}, Ann and {Ichinohe}, Yuto and {Ishida}, Manabu and {Ishikawa}, Kumi and {Ishisaki}, Yoshitaka and {Kaastra}, Jelle and {Kallman}, Timothy and {Kara}, Erin and {Katsuda}, Satoru and {Kanemaru}, Yoshiaki and {Kelley}, Richard and {Kilbourne}, Caroline and {Kitamoto}, Shunji and {Kobayashi}, Shogo and {Kohmura}, Takayoshi and {Kubota}, Aya and {Leutenegger}, Maurice and {Loewenstein}, Michael and {Maeda}, Yoshitomo and {Markevitch}, Maxim and {Matsumoto}, Hironori and {Matsushita}, Kyoko and {McCammon}, Dan and {McNamara}, Brian and {Mernier}, Fran{\c{c}}ois and {Miller}, Eric D. and {Miller}, Jon M. and {Mitsuishi}, Ikuyuki and {Mizumoto}, Misaki and {Mizuno}, Tsunefumi and {Mori}, Koji and {Mukai}, Koji and {Murakami}, Hiroshi and {Mushotzky}, Richard and {Nakajima}, Hiroshi and {Nakazawa}, Kazuhiro and {Ness}, Jan-Uwe and {Nobukawa}, Kumiko and {Nobukawa}, Masayoshi and {Noda}, Hirofumi and {Odaka}, Hirokazu and {Ogawa}, Shoji and {Ogorzalek}, Anna and {Okajima}, Takashi and {Ota}, Naomi and {Paltani}, Stephane and {Petre}, Robert and {Plucinsky}, Paul and {Porter}, Frederick Scott and {Pottschmidt}, Katja and {Sato}, Kosuke and {Sato}, Toshiki and {Sawada}, Makoto and {Seta}, Hiromi and {Shidatsu}, Megumi and {Simionescu}, Aurora and {Smith}, Randall and {Suzuki}, Hiromasa and {Szymkowiak}, Andrew and {Takahashi}, Hiromitsu and {Takeo}, Mai and {Tamagawa}, Toru and {Tamura}, Keisuke and {Tanaka}, Takaaki and {Tanimoto}, Atsushi and {Tashiro}, Makoto and {Terada}, Yukikatsu and {Terashima}, Yuichi and {Tsuboi}, Yohko and {Tsujimoto}, Masahiro and {Tsunemi}, Hiroshi and {Tsuru}, Takeshi G. and {Uchida}, Hiroyuki and {Uchida}, Nagomi and {Uchida}, Yuusuke and {Uchiyama}, Hideki and {Ueda}, Yoshihiro and {Uno}, Shinichiro and {Vink}, Jacco and {Watanabe}, Shin and {Williams}, Brian J. and {Yamada}, Satoshi and {Yamada}, Shinya and {Yamaguchi}, Hiroya and {Yamaoka}, Kazutaka and {Yamasaki}, Noriko and {Yamauchi}, Makoto and {Yamauchi}, Shigeo and {Yaqoob}, Tahir and {Yoneyama}, Tomokage and {Yoshida}, Tessei and {Yukita}, Mihoko and {Zhuravleva}, Irina and {Braito}, Valentina and {Cond{\`o}}, Pierpaolo and {Fukumura}, Keigo and {Gonzalez}, Adam and {Luminari}, Alfredo and {Miyamoto}, Aiko and {Mizukawa}, Ryuki and {Reeves}, James and {Sato}, Riki and {Tombesi}, Francesco and {Xu}, Yerong},
        title = "{Structured ionized winds shooting out from a quasar at relativistic speeds}",
      journal = {\nat},
         year = 2025,
        month = may,
       volume = {641},
       number = {8065},
        pages = {1132-1136},
          doi = {10.1038/s41586-025-08968-2},
archivePrefix = {arXiv},
       eprint = {2505.09171},
 primaryClass = {astro-ph.HE},
       adsurl = {https://ui.adsabs.harvard.edu/abs/2025Natur.641.1132X}
}

@ARTICLE{Peca25,
       author = {{Peca}, Alessandro and {Koss}, Michael J. and {Serafinelli}, Roberto and {Ricci}, Claudio and {Urry}, C. Megan and {Cerini}, Giulia and {Boorman}, Peter G.},
        title = "{NuSTAR Detection of an Absorption Feature in ESP 39607: Evidence for an Ultrafast Inflow?}",
      journal = {\apj},
         year = 2025,
        month = aug,
       volume = {989},
       number = {1},
          eid = {84},
        pages = {84},
          doi = {10.3847/1538-4357/adea4a},
archivePrefix = {arXiv},
       eprint = {2505.07963},
 primaryClass = {astro-ph.HE},
       adsurl = {https://ui.adsabs.harvard.edu/abs/2025ApJ...989...84P}
}

@ARTICLE{Dadina05,
       author = {{Dadina}, M. and {Cappi}, M. and {Malaguti}, G. and {Ponti}, G. and {de Rosa}, A.},
        title = "{X-ray absorption lines suggest matter infalling onto the central black-hole of Mrk 509}",
      journal = {\aap},
         year = 2005,
        month = nov,
       volume = {442},
       number = {2},
        pages = {461-468},
          doi = {10.1051/0004-6361:20042487},
archivePrefix = {arXiv},
       eprint = {astro-ph/0506697},
 primaryClass = {astro-ph},
       adsurl = {https://ui.adsabs.harvard.edu/abs/2005A&A...442..461D}
}

@ARTICLE{YaqoobSerlemitsos05,
       author = {{Yaqoob}, Tahir and {Serlemitsos}, Peter},
        title = "{Iron K Features in the Quasar E1821+643: Evidence for Gravitationally Redshifted Absorption?}",
      journal = {\apj},
         year = 2005,
        month = apr,
       volume = {623},
       number = {1},
        pages = {112-122},
          doi = {10.1086/428432},
archivePrefix = {arXiv},
       eprint = {astro-ph/0502128},
 primaryClass = {astro-ph},
       adsurl = {https://ui.adsabs.harvard.edu/abs/2005ApJ...623..112Y}
}

@INPROCEEDINGS{Longinotti07,
       author = {{Longinotti}, A.~L. and {Sim}, S. and {Nandra}, K. and {Cappi}, M. and {O'Neill}, P.},
        title = "{Relativistic Inflow in the Seyfert 1 Mrk 335 Revealed through X-ray Absorption}",
    booktitle = {The Central Engine of Active Galactic Nuclei},
         year = 2007,
       editor = {{Ho}, L.~C. and {Wang}, J. -W.},
       series = {Astronomical Society of the Pacific Conference Series},
       volume = {373},
        month = oct,
        pages = {341},
          doi = {10.48550/arXiv.astro-ph/0612315},
archivePrefix = {arXiv},
       eprint = {astro-ph/0612315},
 primaryClass = {astro-ph},
       adsurl = {https://ui.adsabs.harvard.edu/abs/2007ASPC..373..341L}
}

@ARTICLE{Giustini17,
       author = {{Giustini}, M. and {Costantini}, E. and {De Marco}, B. and {Svoboda}, J. and {Motta}, S.~E. and {Proga}, D. and {Saxton}, R. and {Ferrigno}, C. and {Longinotti}, A.~L. and {Miniutti}, G. and {Grupe}, D. and {Mathur}, S. and {Shappee}, B.~J. and {Prieto}, J.~L. and {Stanek}, K.},
        title = "{Direct probe of the inner accretion flow around the supermassive black hole in NGC 2617}",
      journal = {\aap},
         year = 2017,
        month = jan,
       volume = {597},
          eid = {A66},
        pages = {A66},
          doi = {10.1051/0004-6361/201628686},
archivePrefix = {arXiv},
       eprint = {1608.00233},
 primaryClass = {astro-ph.GA},
       adsurl = {https://ui.adsabs.harvard.edu/abs/2017A&A...597A..66G}
}

@ARTICLE{PoundsPage24,
       author = {{Pounds}, Ken and {Page}, Kim},
        title = "{Low-redshift absorption in the Seyfert galaxy PG1211+143 - a distant inflow maintaining off-plane accretion or the gravitational redshift of matter orbiting the SMBH?}",
      journal = {\mnras},
         year = 2024,
        month = jul,
       volume = {531},
       number = {4},
        pages = {4852-4856},
          doi = {10.1093/mnras/stae1491},
archivePrefix = {arXiv},
       eprint = {2311.09853},
 primaryClass = {astro-ph.HE},
       adsurl = {https://ui.adsabs.harvard.edu/abs/2024MNRAS.531.4852P}
}

@ARTICLE{Pounds18,
       author = {{Pounds}, K.~A. and {Nixon}, C.~J. and {Lobban}, A. and {King}, A.~R.},
        title = "{An ultrafast inflow in the luminous Seyfert PG1211+143}",
      journal = {\mnras},
         year = 2018,
        month = dec,
       volume = {481},
       number = {2},
        pages = {1832-1838},
          doi = {10.1093/mnras/sty2359},
archivePrefix = {arXiv},
       eprint = {1808.09373},
 primaryClass = {astro-ph.HE},
       adsurl = {https://ui.adsabs.harvard.edu/abs/2018MNRAS.481.1832P}
}

@ARTICLE{Reeves05,
       author = {{Reeves}, J.~N. and {Pounds}, K. and {Uttley}, P. and {Kraemer}, S. and {Mushotzky}, R. and {Yaqoob}, T. and {George}, I.~M. and {Turner}, T.~J.},
        title = "{Evidence for Gravitational Infall of Matter onto the Supermassive Black Hole in the Quasar PG 1211+143?}",
      journal = {\apjl},
         year = 2005,
        month = nov,
       volume = {633},
       number = {2},
        pages = {L81-L84},
          doi = {10.1086/498568},
archivePrefix = {arXiv},
       eprint = {astro-ph/0509280},
 primaryClass = {astro-ph},
       adsurl = {https://ui.adsabs.harvard.edu/abs/2005ApJ...633L..81R}
}

@ARTICLE{Miller14,
       author = {{Miller}, J.~M. and {Raymond}, J. and {Kallman}, T.~R. and {Maitra}, D. and {Fabian}, A.~C. and {Proga}, D. and {Reynolds}, C.~S. and {Reynolds}, M.~T. and {Degenaar}, N. and {King}, A.~L. and {Cackett}, E.~M. and {Kennea}, J.~A. and {Beardmore}, A.},
        title = "{Chandra Spectroscopy of MAXI J1305-704: Detection of an Infalling Black Hole Disk Wind?}",
      journal = {\apj},
         year = 2014,
        month = jun,
       volume = {788},
       number = {1},
          eid = {53},
        pages = {53},
          doi = {10.1088/0004-637X/788/1/53},
archivePrefix = {arXiv},
       eprint = {1306.2915},
 primaryClass = {astro-ph.HE},
       adsurl = {https://ui.adsabs.harvard.edu/abs/2014ApJ...788...53M}
}

@ARTICLE{Trueba20,
       author = {{Trueba}, Nicolas and {Miller}, J.~M. and {Fabian}, A.~C. and {Kaastra}, J. and {Kallman}, T. and {Lohfink}, A. and {Proga}, D. and {Raymond}, J. and {Reynolds}, C. and {Reynolds}, M. and {Zoghbi}, A.},
        title = "{A Redshifted Inner Disk Atmosphere and Transient Absorbers in the Ultracompact Neutron Star X-Ray Binary 4U 1916-053}",
      journal = {\apjl},
         year = 2020,
        month = aug,
       volume = {899},
       number = {1},
          eid = {L16},
        pages = {L16},
          doi = {10.3847/2041-8213/aba9de},
archivePrefix = {arXiv},
       eprint = {2008.01083},
 primaryClass = {astro-ph.HE},
       adsurl = {https://ui.adsabs.harvard.edu/abs/2020ApJ...899L..16T}
}

@ARTICLE{Proga04,
       author = {{Proga}, Daniel and {Kallman}, Timothy R.},
        title = "{Dynamics of Line-driven Disk Winds in Active Galactic Nuclei. II. Effects of Disk Radiation}",
      journal = {\apj},
         year = 2004,
        month = dec,
       volume = {616},
       number = {2},
        pages = {688-695},
          doi = {10.1086/425117},
archivePrefix = {arXiv},
       eprint = {astro-ph/0408293},
 primaryClass = {astro-ph},
       adsurl = {https://ui.adsabs.harvard.edu/abs/2004ApJ...616..688P}
}

@ARTICLE{Proga00,
       author = {{Proga}, Daniel and {Stone}, James M. and {Kallman}, Timothy R.},
        title = "{Dynamics of Line-driven Disk Winds in Active Galactic Nuclei}",
      journal = {\apj},
         year = 2000,
        month = nov,
       volume = {543},
       number = {2},
        pages = {686-696},
          doi = {10.1086/317154},
archivePrefix = {arXiv},
       eprint = {astro-ph/0005315},
 primaryClass = {astro-ph},
       adsurl = {https://ui.adsabs.harvard.edu/abs/2000ApJ...543..686P}
}

@ARTICLE{TomimatsuTakahashi01,
       author = {{Tomimatsu}, Akira and {Takahashi}, Masaaki},
        title = "{Black Hole Magnetospheres around Thin Disks Driving Inward and Outward Winds}",
      journal = {\apj},
         year = 2001,
        month = may,
       volume = {552},
       number = {2},
        pages = {710-717},
          doi = {10.1086/320575},
archivePrefix = {arXiv},
       eprint = {astro-ph/0011185},
 primaryClass = {astro-ph},
       adsurl = {https://ui.adsabs.harvard.edu/abs/2001ApJ...552..710T}
}

@ARTICLE{Hirose04,
       author = {{Hirose}, Shigenobu and {Krolik}, Julian H. and {De Villiers}, Jean-Pierre and {Hawley}, John F.},
        title = "{Magnetically Driven Accretion Flows in the Kerr Metric. II. Structure of the Magnetic Field}",
      journal = {\apj},
         year = 2004,
        month = may,
       volume = {606},
       number = {2},
        pages = {1083-1097},
          doi = {10.1086/383184},
archivePrefix = {arXiv},
       eprint = {astro-ph/0311500},
 primaryClass = {astro-ph},
       adsurl = {https://ui.adsabs.harvard.edu/abs/2004ApJ...606.1083H}
}

@ARTICLE{HawleyKrolik06,
       author = {{Hawley}, John F. and {Krolik}, Julian H.},
        title = "{Magnetically Driven Jets in the Kerr Metric}",
      journal = {\apj},
         year = 2006,
        month = apr,
       volume = {641},
       number = {1},
        pages = {103-116},
          doi = {10.1086/500385},
archivePrefix = {arXiv},
       eprint = {astro-ph/0512227},
 primaryClass = {astro-ph},
       adsurl = {https://ui.adsabs.harvard.edu/abs/2006ApJ...641..103H}
}

@ARTICLE{Fukumura07,
       author = {{Fukumura}, Keigo and {Takahashi}, Masaaki and {Tsuruta}, Sachiko},
        title = "{Magnetohydrodynamic Shocks in Nonequatorial Plasma Flows around a Black Hole}",
      journal = {\apj},
         year = 2007,
        month = mar,
       volume = {657},
       number = {1},
        pages = {415-427},
          doi = {10.1086/510660},
archivePrefix = {arXiv},
       eprint = {astro-ph/0602568},
 primaryClass = {astro-ph},
       adsurl = {https://ui.adsabs.harvard.edu/abs/2007ApJ...657..415F}
}

@ARTICLE{Fukumura04,
       author = {{Fukumura}, Keigo and {Tsuruta}, Sachiko},
        title = "{Isothermal Shock Formation in Nonequatorial Accretion Flows around Kerr Black Holes}",
      journal = {\apj},
         year = 2004,
        month = aug,
       volume = {611},
       number = {2},
        pages = {964-976},
          doi = {10.1086/422243},
archivePrefix = {arXiv},
       eprint = {astro-ph/0405269},
 primaryClass = {astro-ph},
       adsurl = {https://ui.adsabs.harvard.edu/abs/2004ApJ...611..964F}
}

@ARTICLE{Punsly09,
       author = {{Punsly}, Brian and {Igumenshchev}, Igor V. and {Hirose}, Shigenobu},
        title = "{Three-Dimensional Simulations of Vertical Magnetic Flux in the Immediate Vicinity of Black Holes}",
      journal = {\apj},
         year = 2009,
        month = oct,
       volume = {704},
       number = {2},
        pages = {1065-1085},
          doi = {10.1088/0004-637X/704/2/1065},
archivePrefix = {arXiv},
       eprint = {0908.3697},
 primaryClass = {astro-ph.HE},
       adsurl = {https://ui.adsabs.harvard.edu/abs/2009ApJ...704.1065P}
}

@ARTICLE{Ripperda22,
       author = {{Ripperda}, B. and {Liska}, M. and {Chatterjee}, K. and {Musoke}, G. and {Philippov}, A.~A. and {Markoff}, S.~B. and {Tchekhovskoy}, A. and {Younsi}, Z.},
        title = "{Black Hole Flares: Ejection of Accreted Magnetic Flux through 3D Plasmoid-mediated Reconnection}",
      journal = {\apjl},
         year = 2022,
        month = jan,
       volume = {924},
       number = {2},
          eid = {L32},
        pages = {L32},
          doi = {10.3847/2041-8213/ac46a1},
archivePrefix = {arXiv},
       eprint = {2109.15115},
 primaryClass = {astro-ph.HE},
       adsurl = {https://ui.adsabs.harvard.edu/abs/2022ApJ...924L..32R}
}

@ARTICLE{Endo25,
       author = {{Endo}, Yota and {Ishihara}, Hideki and {Takahashi}, Masaaki},
        title = "{Vacuum magnetospheres around Kerr black holes with a thin disk}",
      journal = {\prd},
         year = 2025,
        month = jun,
       volume = {111},
       number = {12},
          eid = {123025},
        pages = {123025},
          doi = {10.1103/PhysRevD.111.123025},
archivePrefix = {arXiv},
       eprint = {2501.00474},
 primaryClass = {gr-qc},
       adsurl = {https://ui.adsabs.harvard.edu/abs/2025PhRvD.111l3025E}
}

@ARTICLE{Lu95,
       author = {{Lu}, J. -F. and {Yu}, K.~N. and {Young}, E.~C.~M.},
        title = "{On the Keplerian circular motion of relativistic fluids.}",
      journal = {\aap},
         year = 1995,
        month = dec,
       volume = {304},
        pages = {662},
       adsurl = {https://ui.adsabs.harvard.edu/abs/1995A&A...304..662L}
}

@ARTICLE{Chakrabarti89,
       author = {{Chakrabarti}, Sandip K.},
        title = "{Standing Rankine-Hugoniot Shocks in the Hybrid Model Flows of the Black Hole Accretion and Winds}",
      journal = {\apj},
         year = 1989,
        month = dec,
       volume = {347},
        pages = {365},
          doi = {10.1086/168125},
       adsurl = {https://ui.adsabs.harvard.edu/abs/1989ApJ...347..365C}
}

@BOOK{Chakrabarti90,
       author = {{Chakrabarti}, Sandip K.},
        title = "{Theory of Transonic Astrophysical Flows}",
         year = 1990,
          doi = {10.1142/1091},
       adsurl = {https://ui.adsabs.harvard.edu/abs/1990ttaf.book.....C}
}

@ARTICLE{Manmoto00,
       author = {{Manmoto}, T.},
        title = "{Advection-dominated Accretion Flow around a Kerr Black Hole}",
      journal = {\apj},
         year = 2000,
        month = may,
       volume = {534},
       number = {2},
        pages = {734-746},
          doi = {10.1086/308768},
       adsurl = {https://ui.adsabs.harvard.edu/abs/2000ApJ...534..734M}
}

@ARTICLE{Ghisellini04,
       author = {{Ghisellini}, G. and {Haardt}, F. and {Matt}, G.},
        title = "{Aborted jets and the X-ray emission of radio-quiet AGNs}",
      journal = {\aap},
         year = 2004,
        month = jan,
       volume = {413},
        pages = {535-545},
          doi = {10.1051/0004-6361:20031562},
archivePrefix = {arXiv},
       eprint = {astro-ph/0310106},
 primaryClass = {astro-ph},
       adsurl = {https://ui.adsabs.harvard.edu/abs/2004A&A...413..535G}
}

@ARTICLE{Takahashi02,
       author = {{Takahashi}, Masaaki},
        title = "{Transmagnetosonic Accretion in a Black Hole Magnetosphere}",
      journal = {\apj},
         year = 2002,
        month = may,
       volume = {570},
       number = {1},
        pages = {264-276},
          doi = {10.1086/339497},
archivePrefix = {arXiv},
       eprint = {astro-ph/0201327},
 primaryClass = {astro-ph},
       adsurl = {https://ui.adsabs.harvard.edu/abs/2002ApJ...570..264T}
}

@ARTICLE{Takahashi08,
       author = {{Takahashi}, Masaaki and {Tomimatsu}, Akira},
        title = "{Constraints on the evolution of black hole spin due to magnetohydrodynamic accretion}",
      journal = {\prd},
         year = 2008,
        month = jul,
       volume = {78},
       number = {2},
          eid = {023012},
        pages = {023012},
          doi = {10.1103/PhysRevD.78.023012},
archivePrefix = {arXiv},
       eprint = {0805.4287},
 primaryClass = {astro-ph},
       adsurl = {https://ui.adsabs.harvard.edu/abs/2008PhRvD..78b3012T}
}

@ARTICLE{Crenshaw03,
       author = {{Crenshaw}, D. Michael and {Kraemer}, Steven B. and {George}, Ian M.},
        title = "{Mass Loss from the Nuclei of Active Galaxies}",
      journal = {\araa},
         year = 2003,
        month = jan,
       volume = {41},
        pages = {117-167},
          doi = {10.1146/annurev.astro.41.082801.100328},
       adsurl = {https://ui.adsabs.harvard.edu/abs/2003ARA&A..41..117C}
}

@ARTICLE{Blustin05,
       author = {{Blustin}, A.~J. and {Page}, M.~J. and {Fuerst}, S.~V. and {Branduardi-Raymont}, G. and {Ashton}, C.~E.},
        title = "{The nature and origin of Seyfert warm absorbers}",
      journal = {\aap},
         year = 2005,
        month = feb,
       volume = {431},
        pages = {111-125},
          doi = {10.1051/0004-6361:20041775},
archivePrefix = {arXiv},
       eprint = {astro-ph/0411297},
 primaryClass = {astro-ph},
       adsurl = {https://ui.adsabs.harvard.edu/abs/2005A&A...431..111B}
}

@ARTICLE{Steenbrugge05,
       author = {{Steenbrugge}, K.~C. and {Kaastra}, J.~S. and {Crenshaw}, D.~M. and {Kraemer}, S.~B. and {Arav}, N. and {George}, I.~M. and {Liedahl}, D.~A. and {van der Meer}, R.~L.~J. and {Paerels}, F.~B.~S. and {Turner}, T.~J. and {Yaqoob}, T.},
        title = "{Simultaneous X-ray and UV spectroscopy of the Seyfert galaxy NGC 5548. II. Physical conditions in the X-ray absorber}",
      journal = {\aap},
         year = 2005,
        month = may,
       volume = {434},
       number = {2},
        pages = {569-584},
          doi = {10.1051/0004-6361:20047138},
archivePrefix = {arXiv},
       eprint = {astro-ph/0501122},
 primaryClass = {astro-ph},
       adsurl = {https://ui.adsabs.harvard.edu/abs/2005A&A...434..569S}
}

@ARTICLE{McKernan07,
       author = {{McKernan}, B. and {Yaqoob}, T. and {Reynolds}, C.~S.},
        title = "{A soft X-ray study of type I active galactic nuclei observed with Chandra high-energy transmission grating spectrometer}",
      journal = {\mnras},
         year = 2007,
        month = aug,
       volume = {379},
       number = {4},
        pages = {1359-1372},
          doi = {10.1111/j.1365-2966.2007.11993.x},
archivePrefix = {arXiv},
       eprint = {0705.2542},
 primaryClass = {astro-ph},
       adsurl = {https://ui.adsabs.harvard.edu/abs/2007MNRAS.379.1359M}
}

@ARTICLE{Tombesi10,
       author = {{Tombesi}, F. and {Cappi}, M. and {Reeves}, J.~N. and {Palumbo}, G.~G.~C. and {Yaqoob}, T. and {Braito}, V. and {Dadina}, M.},
        title = "{Evidence for ultra-fast outflows in radio-quiet AGNs. I. Detection and statistical incidence of Fe K-shell absorption lines}",
      journal = {\aap},
         year = 2010,
        month = oct,
       volume = {521},
          eid = {A57},
        pages = {A57},
          doi = {10.1051/0004-6361/200913440},
archivePrefix = {arXiv},
       eprint = {1006.2858},
 primaryClass = {astro-ph.HE},
       adsurl = {https://ui.adsabs.harvard.edu/abs/2010A&A...521A..57T}
}

@ARTICLE{Nardini15,
       author = {{Nardini}, E. and {Reeves}, J.~N. and {Gofford}, J. and {Harrison}, F.~A. and {Risaliti}, G. and {Braito}, V. and {Costa}, M.~T. and {Matzeu}, G.~A. and {Walton}, D.~J. and {Behar}, E. and {Boggs}, S.~E. and {Christensen}, F.~E. and {Craig}, W.~W. and {Hailey}, C.~J. and {Matt}, G. and {Miller}, J.~M. and {O'Brien}, P.~T. and {Stern}, D. and {Turner}, T.~J. and {Ward}, M.~J.},
        title = "{Black hole feedback in the luminous quasar PDS 456}",
      journal = {Science},
         year = 2015,
        month = feb,
       volume = {347},
       number = {6224},
        pages = {860-863},
          doi = {10.1126/science.1259202},
archivePrefix = {arXiv},
       eprint = {1502.06636},
 primaryClass = {astro-ph.HE},
       adsurl = {https://ui.adsabs.harvard.edu/abs/2015Sci...347..860N}
}

@ARTICLE{Reeves18,
       author = {{Reeves}, J.~N. and {Braito}, V. and {Nardini}, E. and {Lobban}, A.~P. and {Matzeu}, G.~A. and {Costa}, M.~T.},
        title = "{A New Relativistic Component of the Accretion Disk Wind in PDS 456}",
      journal = {\apjl},
         year = 2018,
        month = feb,
       volume = {854},
       number = {1},
          eid = {L8},
        pages = {L8},
          doi = {10.3847/2041-8213/aaaae1},
archivePrefix = {arXiv},
       eprint = {1801.08899},
 primaryClass = {astro-ph.HE},
       adsurl = {https://ui.adsabs.harvard.edu/abs/2018ApJ...854L...8R}
}

@ARTICLE{Chartas09,
       author = {{Chartas}, G. and {Saez}, C. and {Brandt}, W.~N. and {Giustini}, M. and {Garmire}, G.~P.},
        title = "{Confirmation of and Variable Energy Injection by a Near-Relativistic Outflow in APM 08279+5255}",
      journal = {\apj},
         year = 2009,
        month = nov,
       volume = {706},
       number = {1},
        pages = {644-656},
          doi = {10.1088/0004-637X/706/1/644},
archivePrefix = {arXiv},
       eprint = {0910.0021},
 primaryClass = {astro-ph.HE},
       adsurl = {https://ui.adsabs.harvard.edu/abs/2009ApJ...706..644C}
}

@ARTICLE{Parker17,
       author = {{Parker}, Michael L. and {Pinto}, Ciro and {Fabian}, Andrew C. and {Lohfink}, Anne and {Buisson}, Douglas J.~K. and {Alston}, William N. and {Kara}, Erin and {Cackett}, Edward M. and {Chiang}, Chia-Ying and {Dauser}, Thomas and {De Marco}, Barbara and {Gallo}, Luigi C. and {Garcia}, Javier and {Harrison}, Fiona A. and {King}, Ashley L. and {Middleton}, Matthew J. and {Miller}, Jon M. and {Miniutti}, Giovanni and {Reynolds}, Christopher S. and {Uttley}, Phil and {Vasudevan}, Ranjan and {Walton}, Dominic J. and {Wilkins}, Daniel R. and {Zoghbi}, Abderahmen},
        title = "{The response of relativistic outflowing gas to the inner accretion disk of a black hole}",
      journal = {\nat},
         year = 2017,
        month = mar,
       volume = {543},
       number = {7643},
        pages = {83-86},
          doi = {10.1038/nature21385},
archivePrefix = {arXiv},
       eprint = {1703.00071},
 primaryClass = {astro-ph.HE},
       adsurl = {https://ui.adsabs.harvard.edu/abs/2017Natur.543...83P}
}

@ARTICLE{Pounds03,
       author = {{Pounds}, K.~A. and {Reeves}, J.~N. and {King}, A.~R. and {Page}, K.~L. and {O'Brien}, P.~T. and {Turner}, M.~J.~L.},
        title = "{A high-velocity ionized outflow and XUV photosphere in the narrow emission line quasar PG1211+143}",
      journal = {\mnras},
         year = 2003,
        month = nov,
       volume = {345},
       number = {3},
        pages = {705-713},
          doi = {10.1046/j.1365-8711.2003.07006.x},
archivePrefix = {arXiv},
       eprint = {astro-ph/0303603},
 primaryClass = {astro-ph},
       adsurl = {https://ui.adsabs.harvard.edu/abs/2003MNRAS.345..705P}
}

@ARTICLE{HopkinsElvis10,
       author = {{Hopkins}, Philip F. and {Elvis}, Martin},
        title = "{Quasar feedback: more bang for your buck}",
      journal = {\mnras},
         year = 2010,
        month = jan,
       volume = {401},
       number = {1},
        pages = {7-14},
          doi = {10.1111/j.1365-2966.2009.15643.x},
archivePrefix = {arXiv},
       eprint = {0904.0649},
 primaryClass = {astro-ph.CO},
       adsurl = {https://ui.adsabs.harvard.edu/abs/2010MNRAS.401....7H}
}

@ARTICLE{F10,
       author = {{Fukumura}, Keigo and {Kazanas}, Demosthenes and {Contopoulos}, Ioannis and {Behar}, Ehud},
        title = "{Magnetohydrodynamic Accretion Disk Winds as X-ray Absorbers in Active Galactic Nuclei}",
      journal = {\apj},
         year = 2010,
        month = may,
       volume = {715},
       number = {1},
        pages = {636-650},
          doi = {10.1088/0004-637X/715/1/636},
archivePrefix = {arXiv},
       eprint = {0910.3001},
 primaryClass = {astro-ph.HE},
       adsurl = {https://ui.adsabs.harvard.edu/abs/2010ApJ...715..636F}
}

@ARTICLE{Abramowicz97,
       author = {{Abramowicz}, M.~A. and {Lanza}, A. and {Percival}, M.~J.},
        title = "{Accretion Disks around Kerr Black Holes: Vertical Equilibrium Revisited}",
      journal = {\apj},
         year = 1997,
        month = apr,
       volume = {479},
       number = {1},
        pages = {179-183},
          doi = {10.1086/303869},
       adsurl = {https://ui.adsabs.harvard.edu/abs/1997ApJ...479..179A}
}

@ARTICLE{Lu98,
       author = {{Lu}, Ju-Fu and {Yuan}, Feng},
        title = "{Global solutions of adiabatic accretion flows with isothermal shocks in Kerr black hole geometry}",
      journal = {\mnras},
         year = 1998,
        month = mar,
       volume = {295},
       number = {1},
        pages = {66-74},
          doi = {10.1046/j.1365-8711.1998.29511215.x},
       adsurl = {https://ui.adsabs.harvard.edu/abs/1998MNRAS.295...66L}
}

@ARTICLE{Nixon12,
       author = {{Nixon}, Chris and {King}, Andrew and {Price}, Daniel and {Frank}, Juhan},
        title = "{Tearing up the Disk: How Black Holes Accrete}",
      journal = {\apjl},
         year = 2012,
        month = oct,
       volume = {757},
       number = {2},
          eid = {L24},
        pages = {L24},
          doi = {10.1088/2041-8205/757/2/L24},
archivePrefix = {arXiv},
       eprint = {1209.1393},
 primaryClass = {astro-ph.HE},
       adsurl = {https://ui.adsabs.harvard.edu/abs/2012ApJ...757L..24N}
}

@ARTICLE{GaspariSadowski17,
       author = {{Gaspari}, Massimo and {S{\c{a}}dowski}, Aleksander},
        title = "{Unifying the Micro and Macro Properties of AGN Feeding and Feedback}",
      journal = {\apj},
         year = 2017,
        month = mar,
       volume = {837},
       number = {2},
          eid = {149},
        pages = {149},
          doi = {10.3847/1538-4357/aa61a3},
archivePrefix = {arXiv},
       eprint = {1701.07030},
 primaryClass = {astro-ph.HE},
       adsurl = {https://ui.adsabs.harvard.edu/abs/2017ApJ...837..149G}
}

@ARTICLE{Kobayashi18,
       author = {{Kobayashi}, Hiroshi and {Ohsuga}, Ken and {Takahashi}, Hiroyuki R. and {Kawashima}, Tomohisa and {Asahina}, Yuta and {Takeuchi}, Shun and {Mineshige}, Shin},
        title = "{Three-dimensional structure of clumpy outflow from supercritical accretion flow onto black holes}",
      journal = {\pasj},
         year = 2018,
        month = mar,
       volume = {70},
       number = {2},
          eid = {22},
        pages = {22},
          doi = {10.1093/pasj/psx157},
archivePrefix = {arXiv},
       eprint = {1802.00567},
 primaryClass = {astro-ph.HE},
       adsurl = {https://ui.adsabs.harvard.edu/abs/2018PASJ...70...22K}
}

@ARTICLE{Ricci17,
       author = {{Ricci}, C. and {Trakhtenbrot}, B. and {Koss}, M.~J. and {Ueda}, Y. and {Delvecchio}, I. and {Treister}, E. and {Schawinski}, K. and {Paltani}, S. and {Oh}, K. and {Lamperti}, I. and {Berney}, S. and {Gandhi}, P. and {Ichikawa}, K. and {Bauer}, F.~E. and {Ho}, L.~C. and {Asmus}, D. and {Beckmann}, V. and {Soldi}, S. and {Balokovi{\'c}}, M. and {Gehrels}, N. and {Markwardt}, C.~B.},
        title = "{BAT AGN Spectroscopic Survey. V. X-Ray Properties of the Swift/BAT 70-month AGN Catalog}",
      journal = {\apjs},
         year = 2017,
        month = dec,
       volume = {233},
       number = {2},
          eid = {17},
        pages = {17},
          doi = {10.3847/1538-4365/aa96ad},
archivePrefix = {arXiv},
       eprint = {1709.03989},
 primaryClass = {astro-ph.HE},
       adsurl = {https://ui.adsabs.harvard.edu/abs/2017ApJS..233...17R}
}

@ARTICLE{HawleyKrolik02,
       author = {{Hawley}, John F. and {Krolik}, Julian H.},
        title = "{High-Resolution Simulations of the Plunging Region in a Pseudo-Newtonian Potential: Dependence on Numerical Resolution and Field Topology}",
      journal = {\apj},
         year = 2002,
        month = feb,
       volume = {566},
       number = {1},
        pages = {164-180},
          doi = {10.1086/338059},
archivePrefix = {arXiv},
       eprint = {astro-ph/0110118},
 primaryClass = {astro-ph},
       adsurl = {https://ui.adsabs.harvard.edu/abs/2002ApJ...566..164H}
}

@ARTICLE{Gandhi22,
       author = {{Gandhi}, P. and {Kawamuro}, T. and {D{\'\i}az Trigo}, M. and {Paice}, J.~A. and {Boorman}, P.~G. and {Cappi}, M. and {Done}, C. and {Fabian}, A.~C. and {Fukumura}, K. and {Garc{\'\i}a}, J.~A. and {Greenwell}, C.~L. and {Guainazzi}, M. and {Makishima}, K. and {Tashiro}, M.~S. and {Tomaru}, R. and {Tombesi}, F. and {Ueda}, Y.},
        title = "{Frontiers in accretion physics at high X-ray spectral resolution}",
      journal = {Nature Astronomy},
         year = 2022,
        month = dec,
       volume = {6},
        pages = {1364-1375},
          doi = {10.1038/s41550-022-01857-y},
archivePrefix = {arXiv},
       eprint = {2209.10576},
 primaryClass = {astro-ph.HE},
       adsurl = {https://ui.adsabs.harvard.edu/abs/2022NatAs...6.1364G}
}

@ARTICLE{Tombesi11,
       author = {{Tombesi}, F. and {Cappi}, M. and {Reeves}, J.~N. and {Palumbo}, G.~G.~C. and {Braito}, V. and {Dadina}, M.},
        title = "{Evidence for Ultra-fast Outflows in Radio-quiet Active Galactic Nuclei. II. Detailed Photoionization Modeling of Fe K-shell Absorption Lines}",
      journal = {\apj},
         year = 2011,
        month = nov,
       volume = {742},
       number = {1},
          eid = {44},
        pages = {44},
          doi = {10.1088/0004-637X/742/1/44},
archivePrefix = {arXiv},
       eprint = {1109.2882},
 primaryClass = {astro-ph.HE},
       adsurl = {https://ui.adsabs.harvard.edu/abs/2011ApJ...742...44T}
}

@ARTICLE{Gofford13,
       author = {{Gofford}, Jason and {Reeves}, James N. and {Tombesi}, Francesco and {Braito}, Valentina and {Turner}, T. Jane and {Miller}, Lance and {Cappi}, Massimo},
        title = "{The Suzaku view of highly ionized outflows in AGN - I. Statistical detection and global absorber properties}",
      journal = {\mnras},
         year = 2013,
        month = mar,
       volume = {430},
       number = {1},
        pages = {60-80},
          doi = {10.1093/mnras/sts481},
archivePrefix = {arXiv},
       eprint = {1211.5810},
 primaryClass = {astro-ph.HE},
       adsurl = {https://ui.adsabs.harvard.edu/abs/2013MNRAS.430...60G}
}

@ARTICLE{Shi25,
       author = {{Shi}, Fangzheng and {Guainazzi}, Matteo and {Wang}, Yijun},
        title = "{Fast inflowing ionized absorber tracing the gas dynamics at sub-parsec scale around Mrk 3}",
      journal = {arXiv e-prints},
         year = 2025,
        month = may,
          eid = {arXiv:2505.17428},
        pages = {arXiv:2505.17428},
          doi = {10.48550/arXiv.2505.17428},
archivePrefix = {arXiv},
       eprint = {2505.17428},
 primaryClass = {astro-ph.HE},
       adsurl = {https://ui.adsabs.harvard.edu/abs/2025arXiv250517428S}
}

@ARTICLE{Matzeu23,
       author = {{Matzeu}, G.~A. and {Brusa}, M. and {Lanzuisi}, G. and {Dadina}, M. and {Bianchi}, S. and {Kriss}, G. and {Mehdipour}, M. and {Nardini}, E. and {Chartas}, G. and {Middei}, R. and {Piconcelli}, E. and {Gianolli}, V. and {Comastri}, A. and {Longinotti}, A.~L. and {Krongold}, Y. and {Ricci}, F. and {Petrucci}, P.~O. and {Tombesi}, F. and {Luminari}, A. and {Zappacosta}, L. and {Miniutti}, G. and {Gaspari}, M. and {Behar}, E. and {Bischetti}, M. and {Mathur}, S. and {Perna}, M. and {Giustini}, M. and {Grandi}, P. and {Torresi}, E. and {Vignali}, C. and {Bruni}, G. and {Cappi}, M. and {Costantini}, E. and {Cresci}, G. and {De Marco}, B. and {De Rosa}, A. and {Gilli}, R. and {Guainazzi}, M. and {Kaastra}, J. and {Kraemer}, S. and {La Franca}, F. and {Marconi}, A. and {Panessa}, F. and {Ponti}, G. and {Proga}, D. and {Ursini}, F. and {Baldini}, P. and {Fiore}, F. and {King}, A.~R. and {Maiolino}, R. and {Matt}, G. and {Merloni}, A.},
        title = "{Supermassive Black Hole Winds in X-rays: SUBWAYS. I. Ultra-fast outflows in quasars beyond the local Universe}",
      journal = {\aap},
         year = 2023,
        month = feb,
       volume = {670},
          eid = {A182},
        pages = {A182},
          doi = {10.1051/0004-6361/202245036},
archivePrefix = {arXiv},
       eprint = {2212.02960},
 primaryClass = {astro-ph.HE},
       adsurl = {https://ui.adsabs.harvard.edu/abs/2023A&A...670A.182M}
}

@ARTICLE{Gianolli24,
       author = {{Gianolli}, V.~E. and {Bianchi}, S. and {Petrucci}, P.-O. and {Brusa}, M. and {Chartas}, G. and {Lanzuisi}, G. and {Matzeu}, G.~A. and {Parra}, M. and {Ursini}, F. and {Behar}, E. and {Bischetti}, M. and {Comastri}, A. and {Costantini}, E. and {Cresci}, G. and {Dadina}, M. and {De Marco}, B. and {De Rosa}, A. and {Fiore}, F. and {Gaspari}, M. and {Gilli}, R. and {Giustini}, M. and {Guainazzi}, M. and {King}, A.~R. and {Kraemer}, S. and {Kriss}, G. and {Krongold}, Y. and {La Franca}, F. and {Longinotti}, A.~L. and {Luminari}, A. and {Maiolino}, R. and {Marconi}, A. and {Mathur}, S. and {Matt}, G. and {Mehdipour}, M. and {Merloni}, A. and {Middei}, R. and {Miniutti}, G. and {Nardini}, E. and {Panessa}, F. and {Perna}, M. and {Piconcelli}, E. and {Ponti}, G. and {Ricci}, F. and {Serafinelli}, R. and {Tombesi}, F. and {Vignali}, C. and {Zappacosta}, L.},
        title = "{Supermassive Black Hole Winds in X-rays: SUBWAYS. III. A population study on ultra-fast outflows}",
      journal = {\aap},
         year = 2024,
        month = jul,
       volume = {687},
          eid = {A235},
        pages = {A235},
          doi = {10.1051/0004-6361/202348908},
archivePrefix = {arXiv},
       eprint = {2403.09538},
 primaryClass = {astro-ph.GA},
       adsurl = {https://ui.adsabs.harvard.edu/abs/2024A&A...687A.235G}
}

@ARTICLE{Serafinelli19,
       author = {{Serafinelli}, Roberto and {Tombesi}, Francesco and {Vagnetti}, Fausto and {Piconcelli}, Enrico and {Gaspari}, Massimo and {Saturni}, Francesco G.},
        title = "{Multiphase quasar-driven outflows in PG 1114+445. I. Entrained ultra-fast outflows}",
      journal = {\aap},
         year = 2019,
        month = jul,
       volume = {627},
          eid = {A121},
        pages = {A121},
          doi = {10.1051/0004-6361/201935275},
archivePrefix = {arXiv},
       eprint = {1906.02765},
 primaryClass = {astro-ph.GA},
       adsurl = {https://ui.adsabs.harvard.edu/abs/2019A&A...627A.121S}
}

@ARTICLE{Yamada24,
       author = {{Yamada}, Satoshi and {Kawamuro}, Taiki and {Mizumoto}, Misaki and {Ricci}, Claudio and {Ogawa}, Shoji and {Noda}, Hirofumi and {Ueda}, Yoshihiro and {Enoto}, Teruaki and {Kokubo}, Mitsuru and {Minezaki}, Takeo and {Sameshima}, Hiroaki and {Horiuchi}, Takashi and {Mizukoshi}, Shoichiro},
        title = "{X-Ray Winds in Nearby-to-distant Galaxies (X-WING). I. Legacy Surveys of Galaxies with Ultrafast Outflows and Warm Absorbers in z {\ensuremath{\sim}} 0{\textendash}4}",
      journal = {\apjs},
         year = 2024,
        month = sep,
       volume = {274},
       number = {1},
          eid = {8},
        pages = {8},
          doi = {10.3847/1538-4365/ad5961},
archivePrefix = {arXiv},
       eprint = {2405.02391},
 primaryClass = {astro-ph.HE},
       adsurl = {https://ui.adsabs.harvard.edu/abs/2024ApJS..274....8Y}
}
\bibliographystyle{aasjournalv7}

\end{document}